\documentclass[lettersize,journal]{IEEEtran}
\usepackage{amsmath,amsfonts}

\usepackage{algorithm}
\usepackage{algorithmicx}
\usepackage{algpseudocode}
\usepackage{colortbl}
\usepackage{anyfontsize}
\usepackage{listings}
\usepackage{array}
\usepackage[table,xcdraw]{xcolor}
\usepackage{stfloats}
\usepackage{url}
\usepackage{verbatim}
\usepackage{graphicx}
\usepackage{booktabs}
\usepackage{bbding}
\usepackage{multirow}
\usepackage[accsupp]{axessibility}
\usepackage{subcaption}
\usepackage{pifont}
\usepackage[colorlinks=true,   
            linkcolor=red,     
            citecolor=blue,    
            urlcolor=magenta,  
            allbordercolors=white]{hyperref} 
\usepackage{orcidlink}
\begin{document}


\title{Hierarchical Spatial-Frequency Aggregation for Spectral Deconvolution Imaging}


\author{Tao Lv\orcidlink{0009-0006-8269-7623}, Daoming Zhou, Chenglong Huang\orcidlink{ 0009-0007-4552-7737}, Chongde Zi, Linsen Chen\orcidlink{0000-0002-1259-135X},  Xun Cao\orcidlink{0000-0003-3094-4371},~\IEEEmembership{Member, IEEE}

\thanks{Tao Lv, Daoming Zhou, Chenglong Huang, Chongde Zi, Linsen Chen and Xun Cao are with Nanjing University, Nanjing, 210023, China. E-mail: $\lbrace$lvtao, zhou daoming, chenglong-huang$\rbrace$ @smail.nju.edu.cn, $\lbrace$zichongde, chenls,  caoxun$\rbrace$ @nju.edu.cn.}
\thanks{Xun Cao is the corresponding author.}
}

\markboth{Journal of \LaTeX\ Class Files,~Vol.~14, No.~8, August~2021}%
{Shell \MakeLowercase{\textit{et al.}}: A Sample Article Using IEEEtran.cls for IEEE Journals}


\maketitle


\begin{abstract}

Computational spectral imaging (CSI) achieves real-time hyperspectral imaging through co-designed optics and algorithms, 
but typical CSI methods suffer from a bulky footprint and limited fidelity.
Therefore, Spectral Deconvolution imaging (SDI) methods based on PSF engineering have been proposed to achieve high-fidelity compact CSI design recently. However, the composite convolution–integration operations of SDI render the normal-equation coefficient matrix scene-dependent, which hampers the efficient exploitation of imaging priors and poses challenges for accurate reconstruction.
To tackle the inherent data-dependent operators in SDI, we introduce a Hierarchical Spatial–Spectral Aggregation Unfolding Framework (HSFAUF). By decomposing subproblems and projecting them into the frequency domain, HSFAUF transforms nonlinear processes into linear mappings, thereby enabling efficient solutions.
Furthermore, to integrate spatial–spectral priors during iterative refinement, we propose a Spatial–Frequency Aggregation Transformer (SFAT), which explicitly aggregates information across spatial and frequency domains. 
%
By integrating SFAT into HSFAUF, we develop a Transformer-based deep unfolding method, \textbf{H}ierarchical \textbf{S}patial-\textbf{F}requency \textbf{A}ggregation \textbf{U}nfolding \textbf{T}ransformer (HSFAUT), to solve the inverse problem of SDI. Systematic simulated and real experiments show that HSFAUT surpasses SOTA methods with cheaper memory and computational costs, while exhibiting optimal performance on different SDI systems.

\end{abstract}


\begin{figure}[t]
\begin{center}
  \includegraphics[width=1\linewidth]{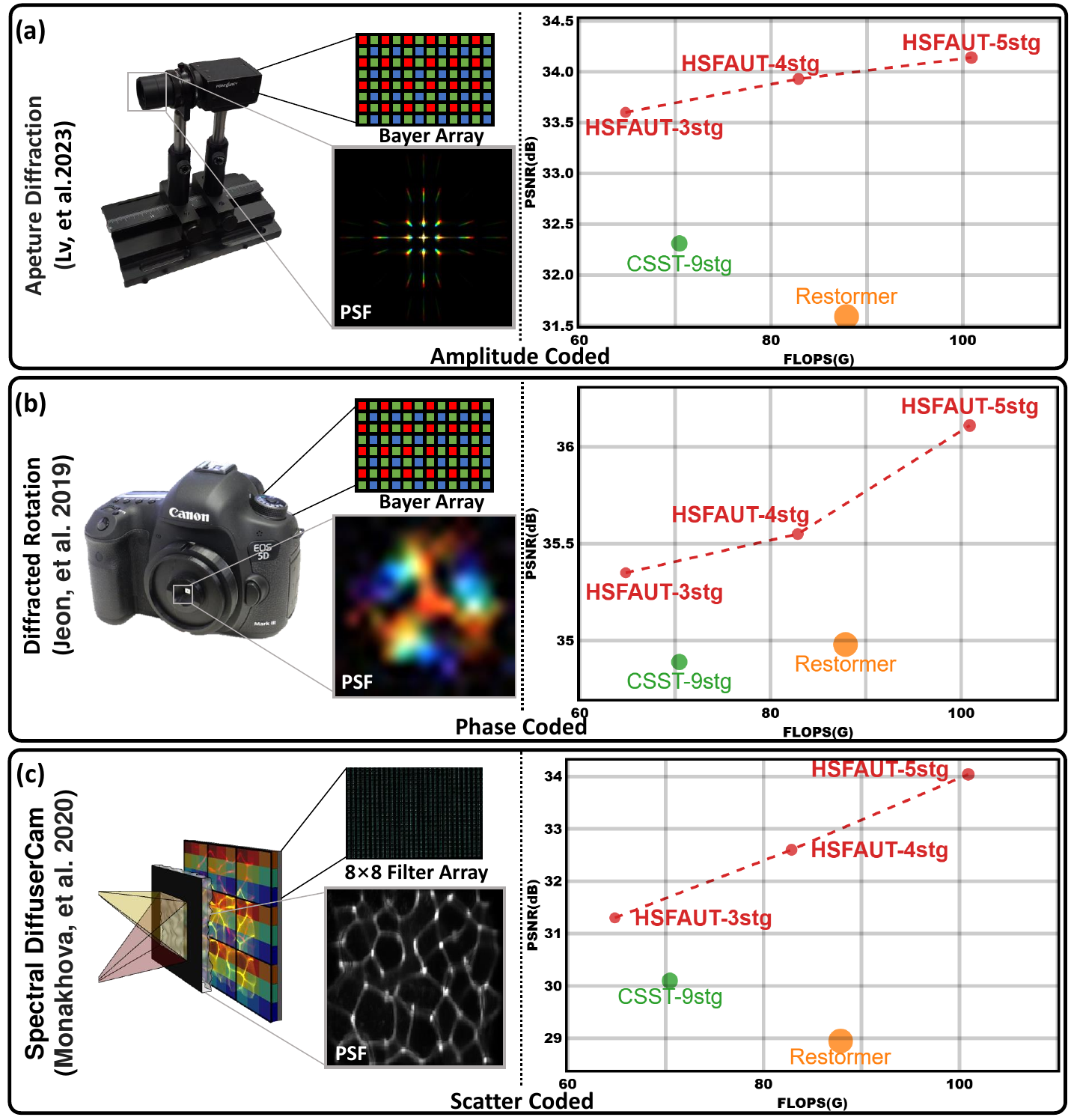}
\end{center}
\vspace{-0.2cm} 
\caption{PSNR-Params-FLOPS comparisons with different reconstruction methods on three typical Spectral Deconvolution Imaging (SDI) systems. The vertical axis is PSNR (in dB performance), the horizontal axis is FLOPS (computational costs), and the circle radius is Params (memory costs).}
\vspace{-0.3cm}
\label{Fig1}
\end{figure}

\section{Introduction}
\label{sec:intro}

\begin{figure*}[t]
\begin{center}
  \includegraphics[width=1\linewidth]{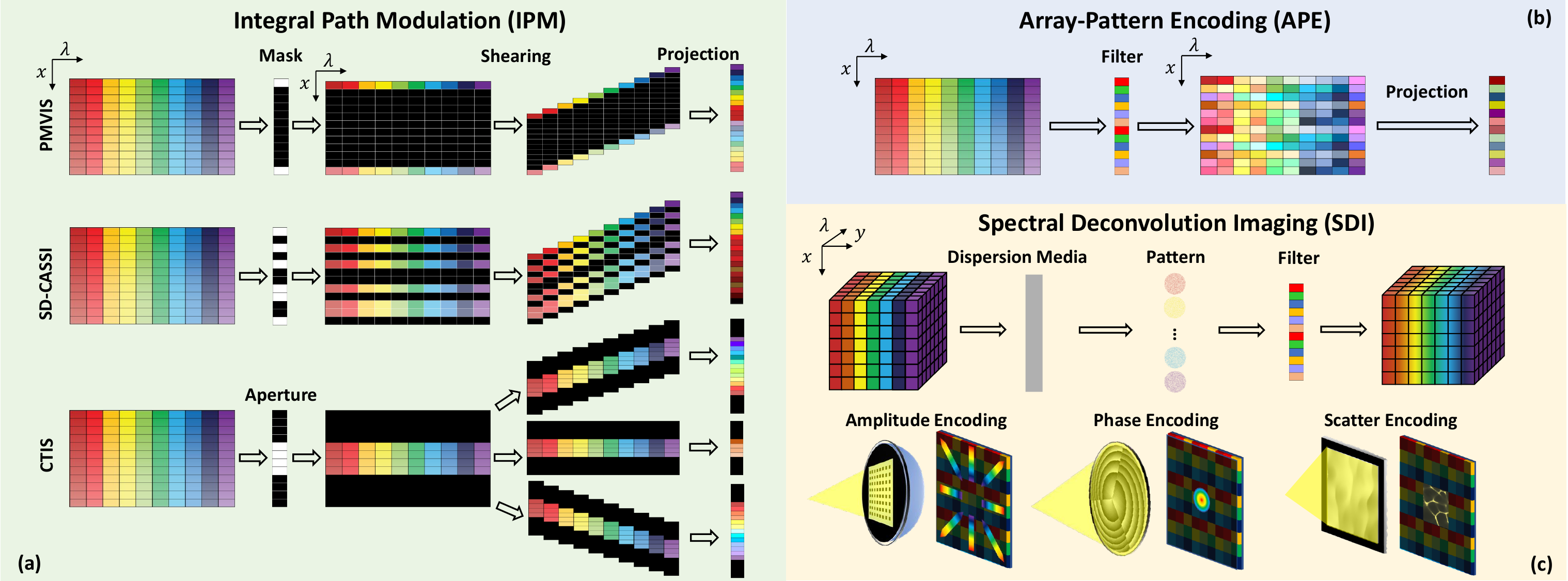}
\end{center}
\vspace{-0.2cm}
\caption{Depicts three paradigms of computational spectral imaging, each based on different principles. SDI offers \textbf{superior integration} and \textbf{fidelity} compared to IPM and APE.} 
\label{fig2}
\vspace{-0.2cm} 
\setlength{\belowcaptionskip}{-0.3cm}
\end{figure*}
Hyperspectral images (HSIs) capture high-resolution spectra at each spatial location, providing a spectral representation that reveals the rich characteristics of various components and materials,
offering a high-dimensional visual capability beyond human vision. Thus, HSIs have found widespread applications in fields such as medical diagnosis~\cite{li2013review}, remote sensing~\cite{yuan2017hyperspectral,goetz1985imaging}, agricultural inspection~\cite{wang2021review}, and machine vision~\cite{bao2023heat}.
However, early hyperspectral imaging techniques were constrained by 2D sensor, requiring spatial or spectral scanning that sacrificed temporal resolution for spectral resolution, restricting their use in dynamic scenes. 
To overcome these challenges, computational spectral imaging (CSI)~\cite{cao2016computational} integrates optics, electronics, and algorithms to enhance imaging capabilities~\cite{SDCASSI,PMVIS,CTIS}. 

CSI systems utilize diverse designs but share a common approach: capturing high-dimensional spectral data via compressed sampling and solving a sparsely constrained inverse problem, making the \textbf{compact structure} and \textbf{high-fidelity reconstruction} critical.
However, while typical CSI systems such as coded aperture snapshot spectral imaging (CASSI) are interpretable, they often suffer from large system footprints, whereas compact CSI architectures based on array-pattern filtering or response encoding rely heavily on data priors for reconstruction. In light of this trade-off, we classify CSI systems according to their structural characteristics and the nature of the associated inverse problems, as illustrated in Figure~\ref{fig2}: \textbf{I}ntegral-\textbf{P}ath \textbf{M}odulation (IPM), \textbf{A}rray-\textbf{P}attern \textbf{E}ncoding (APE), and \textbf{S}pectral \textbf{D}econvolution \textbf{I}maging (SDI):


$\bullet$ \textbf{IPM} employs an occlusion mask or field stop to modulate high-dimensional data, followed by projection through dispersive optics to form measurements~\cite{SDCASSI,DDCASSI,HVIS,PMVIS,CTIS} with complex optical structure and repetitive calibration~\cite{lv2024efficient}. 

$\bullet$ \textbf{APE} replaces dispersive optics by integrating micro- or nano-optical structures with designed spatial patterns onto the sensor, enabling encoded transmittance or  response~\cite{wang2007concept,bian2024broadband,yako2023video,wang2019single,yesilkoy2019ultrasensitive}. APE enables CSI with significantly reduced footprint. However, the reconstruction relies heavily on data-driven priors, limiting its fidelity in unfamiliar scenarios. 

$\bullet$ \textbf{SDI} achieves \textbf{system compactness} and \textbf{high fidelity} by optimizing point spread functions (PSFs)~\cite{ADIS,DOE_Jeon,diffusercam} with specific filtering. In recent years, hot research topics in deep optics have significantly enriched the SDI configuration design and attracted considerable attention~\cite{wang2024non,shi2024split}. However, because each voxel in the HSIs is projected through specific PSFs, the coefficient matrix in the resulting normal equations becomes scene-dependent. This non-stationarity complicates the inverse problem, demanding reconstruction algorithms with strong decoupling capabilities to recover accurate HSIs.

Specifically, the forward model of SDI can be characterized by a linear transformation of the original signal $I$ into the measurement $M$ through a sensing matrix $\bf{\Phi }$. In this formulation, the coefficient matrices of the normal equations ${{\bf{\Phi }}^\top}{\bf{\Phi }}$, exhibits a strong frequency-dependent response tied to the spatial structure of the scene, and cannot be decoupled independently. Due to the non-diagonal structure of these operators, information across different dimensions becomes highly coupled, resulting in an inherently ill-posed inverse problem that significantly complicates reconstruction. The central challenge, therefore, lies in \textbf{effectively extracting and incorporating imaging priors} to develop efficient algorithms—essential for \textbf{achieving compact and high-fidelity computational spectral imaging}.


To effectively leverage imaging priors, deep unfolding methods offer a promising pathway to integrate domain knowledge while maintaining interpretability. However, existing deep unfolding methods are primarily designed for IPM and APE systems, where ${{\bf{\Phi }}^\top}{\bf{\Phi }}$ possesses certain diagonalisation characteristics. These architectures are largely incompatible with SDI reconstruction due to its distinct imaging characteristics. \textbf{Consequently, the potential of combining SDI-specific properties with deep unfolding remains largely untapped.}

A further challenge arises from the fact that SDI systems employ diverse PSF-engineering principles such as phase, amplitude, and scattering encoding, which yield PSFs with fundamentally different energy distributions. Phase encoding generally produces relatively concentrated PSFs, while amplitude and scattering-based methods tend to generate widely dispersed energy patterns. These variations place heightened demands on the \textbf{receptive field size} and \textbf{long-range perception capability} of conventional spatial-domain deep unfolding frameworks and denoisers. \textbf{A central challenge, therefore, is to leverage the underlying commonality among SDI systems in order to achieve consistently high performance across different encoding principles, without incurring excessive computational or memory costs.}

Notably, in SDI, \textbf{PSF engineering can be reformulated in the frequency domain as a one-to-one mapping}. This transformation converts the nonlinear process in reconstruction into a linear mapping, significantly reducing computational complexity and explicitly incorporating physical constraints. Moreover, it alleviates the algorithm’s reliance on large receptive fields (a requirement stemming from the PSF's dispersed projection of information across the sensor), thereby enhancing both the efficiency and performance of the reconstruction.

Motivated by this, we propose a \textbf{H}ierarchical \textbf{S}patial-\textbf{F}requency \textbf{A}ggregation \textbf{U}nfolding \textbf{F}ramework (HSFAUF) based on maximum a posteriori (MAP) theory. HSFAUF employs a hierarchical structure to iteratively solve convolutional subproblems in the frequency domain and filtering subproblems in the spatial domain. It explicitly aggregates spatial–spectral imaging priors by adaptively refining linear projections across complementary computational domains, resulting in an efficient and physics-informed reconstruction. To further enhance cross-domain feature aggregation, we introduce a \textbf{S}patial–\textbf{}Frequency \textbf{A}ggregation \textbf{T}ransformer (SFAT) that jointly computes self-attention in both spatial and frequency, thereby effectively capturing critical cross-domain cues. By integrating SFAT into the HSFAUF, we develop a versatile reconstruction algorithm, \textbf{H}ierarchical \textbf{S}patial-\textbf{F}requency \textbf{A}ggregation \textbf{U}nfolding \textbf{T}ransformer (HSFAUT). As illustrated in Fig. \ref{Fig1}, HSFAUT achieves state-of-the-art reconstruction performance across various SDI system configurations with reduced computational and memory overhead. A unified benchmarking framework, demonstrated in Fig. \ref{Fig3}, facilitates comprehensive comparison of different algorithms under diverse SDI configuration. Extensive simulations and real-world experiments validate the effectiveness of our approach. In summary, we make the following contributions:


$\bullet$  We first classify CSI into three categories based on inverse problem: IPM, APE, and SDI. Due to the inherent limitations of IPM in structural complexity and APE in reconstruction fidelity for unknown scenes, we highlights the necessity of efficient reconstruction of SDI for achieving high-fidelity compact CSI. Furthermore, a unified evaluation framework is proposed for SDI that enables consistent and comparative assessment of various configurations and algorithms.

$\bullet$  We propose a principled MAP-based framework HSFAUF for efficiently extracting prior information and solving SDI inverse problems across different computational domains. The reasons for the failure of deep-unfolding frameworks in SDI are also discussed.

$\bullet$ We propose a novel denoiser SFAT and plug it into HSFAUF to establish HSFAUT. HSFAUT outperforms previous methods by a large margin on different SDI configuration while requiring cheaper computational and memory costs. 

$\bullet$ The HSFAUT is systematically evaluated through comprehensive simulations and real-world experiments, demonstrating its broad applicability and high reconstruction accuracy. The method achieves superior reconstruction performance and produces visually compelling results, showing strong potential to advance high-fidelity compact CSI.


\begin{table*}[t]
\caption{Typical computational spectral imaging methods are classified according to the category of imaging principles.}
\vspace{-0.2cm}
\begin{center}
\renewcommand{\arraystretch}{1.2} 
\setlength{\tabcolsep}{3.2pt}     
\resizebox{2\columnwidth}{!}{
\begin{tabular}{cccccc}
\hline
\multicolumn{6}{c}{Computational Spectral Imaging (CSI)}                                                                                                                                                                          \\ \hline
\multicolumn{2}{c|}{Integral-Path Modulation (IPM, projection engineering)} & \multicolumn{2}{c|}{Array-Filter Encoding (AFE, filtering engineering)} & \multicolumn{2}{c}{Spectral Deconvolution Imaging (SDI, PSF engineering)} \\ \hline
System                    & \multicolumn{1}{c|}{Principle}                  & System                 & \multicolumn{1}{c|}{Principle}                 & System                             & Principle                            \\ \hline
SD-CASSI~\cite{SDCASSI}      & \multicolumn{1}{c|}{Binary occlusion (random), dispersion} & Bian et al.~\cite{bian2024broadband}      & \multicolumn{1}{c|}{Random broadband filtering, photolithography}                & ADIS~\cite{ADIS}                                 & Amplitude coding, PSFs diffuse  \\ 
DD-CASSI~\cite{DDCASSI}    & \multicolumn{1}{c|}{Binary occlusion (random), dispersion}    & Yako et al. ~\cite{yako2023video}     & \multicolumn{1}{c|}{Random broadband filtering, F-P filters}                    & Jeon et al.~\cite{DOE_Jeon}             &  Phase coding, PSFs rotation                                 \\
PMVIS~\cite{PMVIS} & \multicolumn{1}{c|}{Binary occlusion (uniform), dispersion}          & Zhang et al.~\cite{zhang2023handheld}               & \multicolumn{1}{c|}{Random broadband filtering, film mask}    & Spectral DiffuserCam~\cite{diffusercam}        & Scattering coding, specific scattering pattern   \\
CTIS~\cite{CTIS}   & \multicolumn{1}{c|}{Binary occlusion (partial), dispersion}          & Yang et al.~\cite{yang2022ultraspectral}                 & \multicolumn{1}{c|}{broadband response coding, metasurface}               & SCCD~\cite{arguello2021shift}     & Amplitude and phase coding, PSFs variation                                    \\
DCSI~\cite{lin2014dual}   & \multicolumn{1}{c|}{Dynamic binary occlusion, dispersion}      & Wang et al.~\cite{wang2019single}        & \multicolumn{1}{c|}{Random broadband filtering, photonic crystal}                         & Wang et al.~\cite{wang2024non}                                  & phase coding, PSFs variation                                    \\
CCASSI (Correa et al.)~\cite{correa2014compressive}     & \multicolumn{1}{c|}{Color occlusion, dispersion}                          & Zhao et al.~\cite{zhao2019hyperspectral}                      & \multicolumn{1}{c|}{Random broadband filtering, ink printing}                         & Oktem et al.~\cite{oktem2021high}             & Phase coding, PSFs variation                          \\
CCASSI (Arguello et al.)~\cite{arguello2014colored}      & \multicolumn{1}{c|}{Color occulusion, dispersion}                          & Park et al.~\cite{park2024avian}                      & \multicolumn{1}{c|}{broadband response coding, perovskite}                         & Kar et al~\cite{kar2019compressive}            & binary occlusion and phase coding, PSFs variation                                    \\
CCASSI (Rueda et al.)~\cite{rueda2015multi}     & \multicolumn{1}{c|}{Color occulusion, dispersion}     & Tittl ey al.~\cite{tittl2018imaging}      & \multicolumn{1}{c|}{response coding, metasurface}                         & Gundogan et al.~\cite{gundogan2021computational}                                  & Phase coding and filtering, PSFs variation                                    \\ \hline
\end{tabular}

}
\end{center}
\label{tab1}
\vspace{-0.4cm} 
\end{table*}

\vspace{-0.2cm}
\section{Related Work}
\label{sec:relate}

\subsection{Computational Spectral Imaging}

\subsubsection{Integral-Path Modulation, IPM}
\label{subsec:IPM}

A typical IPM method employs an occlusion mask in the focal plane to leverage the scene's inherent spatial sparsity for encoding spatial information, then followed by the dispersive optics to project 3D data cube as 2D measurement. Representastive techniques such as CASSI~\cite{SDCASSI,DDCASSI} and PMVIS~\cite{PMVIS, HVIS}, have been developed based on this approach, as shown in Fig.~\ref{fig2}(a). 
CTIS~\cite{CTIS} leverages spatially constrained dispersive optics to generate multiple integration paths, yielding different integration patterns for spectral recovery.
Some enhanced methods (e.g., CCASI~\cite{cao2016computational, correa2014compressive, arguello2014colored, rueda2015multi,correa2015snapshot}, DCSI~\cite{cao2016computational,lin2014dual}) modify the system architecture to achieve a more randomized sensing matrix or increased light throughput, yet they remain specific modulations along the integration path~\cite{cao2016computational, cao2021hyperspectral}.
Consequently, IPM approaches struggle with compact integration and require repetitive calibration, limiting their practical applicability.

\subsubsection{Array-Pattern Encoding, APE}
\label{subsec:AFC}
In contrast, the APE approach forgoes the complex integration path design enabled by dispersive optics, acquiring spectral data directly via a on-chip transmittance or response encoded array with a compact structure. 
However, As shown in Fig.~\ref{fig2}(b), APE methods rely on learning-based reconstruction to recover spectral information~\cite{arad2022ntire,MST++}, causing them to depend primarily on data priors learned from training datasets rather than the original data captured by the sensor~\cite{fu2024limitations}.
In APE, transmittance or response encoded array design is the primary focus for optimization, and an improved sensing matrix enhances overall imaging performance~\cite{yako2023video}.
Consequently, imaging spectrometers using specialized arrays—such as thin films~\cite{wang2007concept}, photolithographic coatings~\cite{ bian2024broadband}, Fabry-Pérot random filters~\cite{yako2023video}, planar photonic crystals~\cite{wang2019single}, and metasurfaces~\cite{yesilkoy2019ultrasensitive}—have been demonstrated in laboratory settings. Nonetheless, the inherent challenge of guaranteeing recovery fidelity for unknown scenes, coupled with high manufacturing costs and low sample consistency, limits the broader adoption of APE.

\subsubsection{Spectral Deconvolution Imaging, SDI}
\label{subsec:SDI}
As shown in Fig.~\ref{fig2}(c), SDI achieves high fidelity and compact system design through the optimization of the point spread functions (PSFs) engineering with specific filtering. This integrated approach has led to the development of three representative schemes: amplitude encoding~\cite{ADIS}, phase encoding~\cite{DOE_Jeon, wang2024non} and scatter encoding\cite{diffusercam}.
For SDI systems, the heterogeneity of PSFs and the customization of filter functions exacerbate spatial degradation and spectral aliasing, while providing richer coding~\cite{arguello2021shift}. This arises from two factors: (1) unlike IPM and APE, SDI's convolutional coding acts as frequency-domain modulation, thereby achieving joint spatial-frequency modulation; (2) SDI maps 3D voxels to the sensor via PSFs projections in a predetermined pattern, resulting in a one-to-many mapping that requires the reconstruction algorithm to operate over a broader perceptual range with enhanced solving capabilities.
Therefore, some approaches mitigate the severe SDI inverse problem by enhancing reconstruction quality via auxiliary branching, though at the expense of system compactness and usability~\cite{xu2023hyperspectral}. Other works pursue calibration-free, parallel SDI architectures through domain generalization, but reducing the overall system footprint remains difficult~\cite{lv2024efficient}. Thus, leveraging the intrinsic properties of the SDI process and hardware priors while minimizing computational and storage demands is critical for advancing SDI toward practical applications.

Driven by advances in deep optics, the CSI systems are increasingly adopting the SDI paradigm with spectrally variant PSFs for optical encoding~\cite{wang2024non,shi2024split}.
This evolution makes physics-driven SDI reconstruction algorithms (computational decoders) essential in deep optics, \textbf{which provide a more principled understanding of the underlying inverse problem}. Such algorithms not only improve overall imaging performance but also facilitate the co-design of globally optimized optical structures. Therefore, developing resource-efficient and physics-aware SDI reconstruction methods is crucial for achieving high-fidelity and practical CSI.


\subsection{Reconstruction Methods}
\label{subsec:Rec}


\subsubsection{Reconstruction for Different Systems}
In CSI reconstruction, extensive research has focused on imaging systems based on the IPM principle represented by CASSI~\cite{SDCASSI,DDCASSI}.
Various methods have been developed, including iterative techniques that offer unsupervised solutions to IPM but face significant challenges of being computationally intensive and time-consuming. learning-based methods aim to alleviate this problem and significantly improve the reconstruction accuracy~\cite{l-net,HDnet,TSA-net,DGSMP,BIRNAT}. Notably, Transformer-based reconstruction algorithms have further elevated performance levels~\cite{MST,CST,DAUHST,dong2023residual,zhang2024improving}. In APE, the reconstruction process is treated as a super-resolution problem, benefiting from advanced feature extraction and mapping capabilities~\cite{Unet, MIRNet, MPRNet, bian2024broadband, restormer}. 
In contrast, SDI reconstruction has received less attention, especially regarding the integration of hardware priors with the imaging process. Although Jeon et al. \cite{DOE_Jeon} directly transferred the deep unfolding method in the denoising task to the SDI reconstruction, the approach does not fully decouple the SDI imaging process. 

\begin{figure*}[t]
    \begin{center}
    \includegraphics[width=1\linewidth]{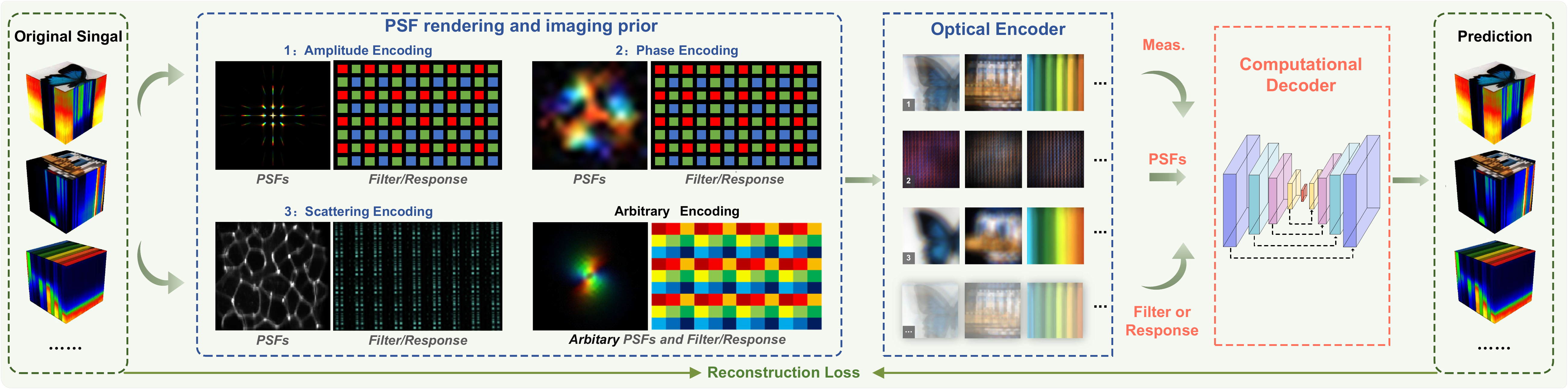}
    \end{center}
\vspace{-0.2cm} 
\caption{Illustration of the unified SDI reconstruction framework. By replacing the PSFs and filter functions customized for different SDI design, this framework facilitates the reconstruction of various SDI configurations.}
\label{Fig3}
\vspace{-0.4cm} 
\end{figure*}

\subsubsection{Deep Unfolding Methods}
Deep unfolding methods, which integrate iterative optimization with deep learning, have been widely and successfully applied to reconstruct data from CASSI systems by effectively leveraging physical imaging priors~\cite{DAUHST, wu2024latent, li2023pixel, dong2023residual}.
In contrast, their application to SDI remains nascent. 
On one hand, these frameworks are designed for IPM, utilizing principles of blocking and dispersion that do not translate directly to SDI~\cite{DAUHST, wu2024latent, li2023pixel}. On the other hand, the denoising priors they incorporate are also grounded in IPM-based mask features, which are either incompatible or ineffective for the SDI inverse problem~\cite{MST,dong2023residual, wang2025sˆ}.
To fully leverage the physical priors and characteristics inherent in SDI, we propose a hierarchical deep unfolding framework that separately explores spatial and frequency domain features, achieving superior results across various SDI systems with reduced computational and memory costs.

\section{Proposed Method}
\label{sec:method}
\subsection{Degradation Model of SDI}
\label{sec:method1}

The primary objective of SDI is to reconstruct the hyperspectral images (HSIs) from convolution-coded and filter-coded measurements. Considering an SDI system uses the PSFs, denoted as $P_\lambda (x,y)$, and a filter function $\Omega _\lambda (x,y)$, the measurement $M(x,y)$ can be expressed as:
\begin{equation}
\setlength{\abovedisplayskip}{0.1cm}
\setlength{\belowdisplayskip}{0.1cm}
    {M(x,y) = \int {{\Omega _\lambda }(x,y) \cdot ({P_\lambda }(x,y)*{I_\lambda }(x,y)})d\lambda } 
\label{3-1-1}
\end{equation}
where $\mathcal{*}$ represents the convolution operation.

By introducing a variable ${J_\lambda }(x,y)$ to denote the intermediate result after convolution, Equation~\ref{3-1-1} can be rewritten:

\begin{equation}
\setlength{\abovedisplayskip}{0.1cm}
\setlength{\belowdisplayskip}{0.1cm}
    \left\{ {\begin{array}{*{20}{c}}
{{J_\lambda }(x,y) = {P_\lambda }(x,y)*{I_\lambda }(x,y)}\\
{M(x,y) = \int {{\Omega _\lambda }(x,y) \cdot {J_\lambda }(x,y)d\lambda } }
\end{array}} \right.
\label{3-1-2}
\end{equation}

The discrete image formation model can be expressed in vector-matrix forms. Let $\bf{I}\in\mathbb{R}^{nC\times1}$ denote the vectorized original HSIs, $\bf{J}\in\mathbb{R}^{nC\times1}$ represent the intermediate iamge after convolution, and $\bf{M}\in\mathbb{R}^{nC\times1}$ represent the captured measurements. Here, $n = H \times W$, while $H,W$ and $C$ denote the height, width, and total number of spectral bands of the original HSIs, respectively. The convolution effect of $P_\lambda (x,y)$ and the integration with filtering ${\Omega _\lambda }(x,y)$ are respectively represented by ${{\bf{\Phi }}_1} \in\mathbb{R}^{nC\times nC}$,${{\bf{\Phi }}_2} \in\mathbb{R}^{nC\times nC}$. Thus, Equation~\ref{3-1-1} can be formulated as:

\begin{equation}
\setlength{\abovedisplayskip}{0cm}
\setlength{\belowdisplayskip}{0cm}
    {\bf{M}} = {{\bf{\Phi }}_2}{{\bf{\Phi }}_1}{\bf{I}} + \bf{n} = {{\bf{\Phi }}_2}{\bf{J}} +\bf{n}
\label{3-1-3}
\end{equation}

\noindent
where ${{\bf{\Phi }}_1}$,${{\bf{\Phi }}_2}$ is a large, sparse, and structured matrix that is challenging to handle directly, $\bf{n}$ is the imaging noise on the measurement, generated by the photon sensing detector.. The hyperspectral reconstruction task for SDI is then formulated as: solving $\bf{I}$ given $\bf{M}$,$\bf{\Phi _1}$,$\bf{\Phi _2}$.











\begin{figure*}[t]
    \begin{center}
    \includegraphics[width=1\linewidth]{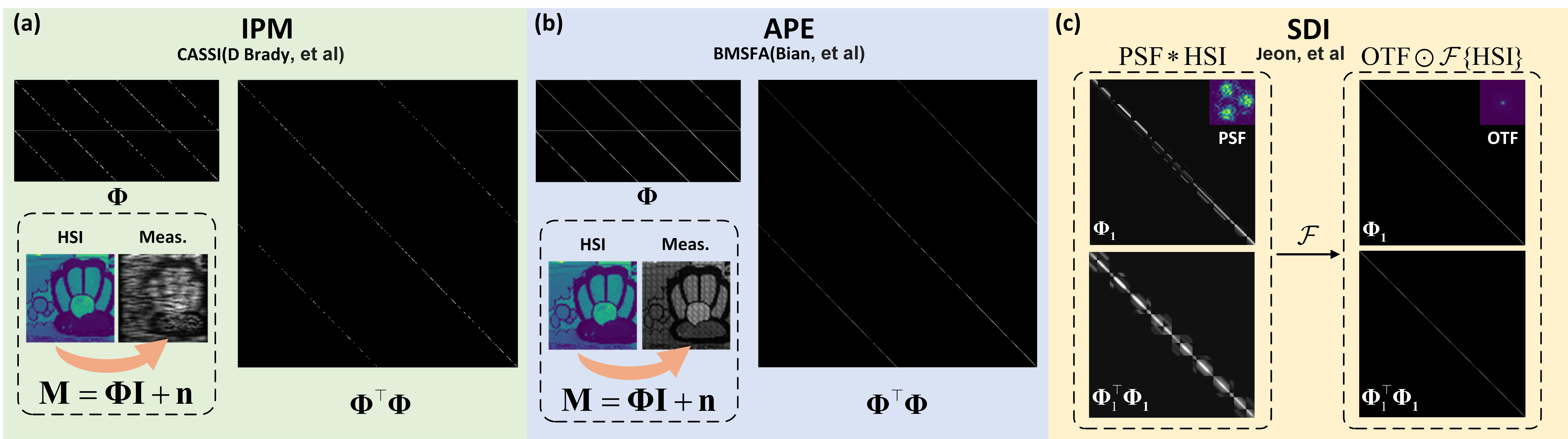}
    \end{center}
\vspace{-0.2cm} 
\caption{Sensing matrices and Hessian matrices of different CSI systems. The inverse process of SDI's PSFs convolution encoding can be transformed in the frequency domain into an optimization process featuring a diagonalised Hessian matrix.}
\vspace{-0.2cm} 
\label{Fig4}
\end{figure*}


\subsection{Optimisation Problem Analysis}
\label{sec:method2}

\subsubsection{Sensing Characteristics of Different CSI Architectures}

To delineate the distinct inverse problem characteristics of SDI compared to IPM and APE, we analyze their respective degradation models. We use CASSI~\cite{SDCASSI} as a representative IPM system and the broadband stochastic encoding scheme proposed by Bian et al.~\cite{bian2024broadband} as a canonical APE implementation.

For CASSI, the original signal ${I_{IPM}} \in {{\mathbb{R}}^{H \times W \times C}}$ is firstly modulated by the physical mask $\mathcal{M} \in {\mathbb{R}^{H \times W \times C}}$ and dispersion optics:
\begin{equation}
\setlength{\abovedisplayskip}{0cm}
\setlength{\belowdisplayskip}{0cm}
    I_{IPM}^{'}(u,v,{n_\lambda }) = ({I_{IPM}} \odot \mathcal{M})(x,y + d({\lambda _n} - {\lambda _c}),{n_\lambda })
\label{3-12-1}
\end{equation}

\noindent
where $I_{IPM}^{'}$ denotes the modulated HSIs, ${n_\lambda } \in [1,2,...,C]$ indexes the spectral channels, and  $\odot$  denotes the element-wise multiplication. 
Then IPM acquires the measurement values by performing band-by-band integration on the modulated signal:
\begin{equation}
\setlength{\abovedisplayskip}{0cm}
\setlength{\belowdisplayskip}{0cm}
    {M}_{IPM} = \sum\limits_{{n_\lambda } = 1}^{{C}} {I_{IPM}^{'}(:,:,{n_\lambda })}  + \bf{n}
\label{3-12-2}
\end{equation}

For the APE method, measurements are obtained directly through band-by-band integration following modulation by a filter or response matrix $\mathcal{Q} \in {R^{H \times W \times C}}$:
\begin{equation}
\setlength{\abovedisplayskip}{0cm}
\setlength{\belowdisplayskip}{0cm}
    {M_{APE}} = \sum\limits_{{n_\lambda } = 1}^{{C}} {({I_{APE}} \odot \mathcal{Q})(:,:,{n_\lambda })} 
\label{3-12-3}
\end{equation}
\noindent
where \(M_{APE} \in \mathbb{R}^{H \times W}\) denotes the measurement from the APE system, and \(I_{APE} \in \mathbb{R}^{H \times W \times C}\) represents the measurement from the APE system.

Equations \ref{3-1-1}, \ref{3-12-2}, and \ref{3-12-3} define the forward imaging processes for different imaging pipelines, which can be uniformly described as ${\bf{M}} = {\bf{\Phi I}} + {\bf{n}}$, where $\bf{\Phi}$ denotes the sensing matrix of different CSI systems. We perform reconstruction by solving the following optimisation problem:
\begin{equation}
\setlength{\abovedisplayskip}{0cm}
\setlength{\belowdisplayskip}{0cm}
    {\bf{\hat I}} = \mathop {\min }\limits_{\bf{I}} \underbrace {\frac{1}{2}{{\left\| {{\bf{M}} - {\bf{\Phi I}}} \right\|}^{\bf{2}}}}_{f({\bf{I}})} + {\bf{\lambda R}}({\bf{I}})
\label{3-12-4}
\end{equation}

The sensing matrix $\mathbf{\Phi}_{SDI}$ exhibits a fundamental structural distinction from those of IPM and APE ($\mathbf{\Phi}_{IPM}$, $\mathbf{\Phi}_{APE}$): it is globally dense rather than block-diagonal. The dense structure contributes to the ill-posedness of SDI reconstruction, which arises from the large and non-uniformly distributed singular values of $\mathbf{\Phi}_{SDI}$. These significant singular values are indicative of substantial information capacity, allowing nearly complete encoding of high-dimensional spectral scenes into the measurements. Consequently, the ill-posedness in SDI can be viewed as a “\textbf{blessing in disguise}”—while it complicates inversion, the difficulty is primarily algorithmic. Through appropriate regularization and learned priors, one can recover high-fidelity information that is embedded but not readily accessible in the measurements. By contrast, the “\textbf{apparent well-posedness}” of IPM and APE systems often results from dimensional reduction, yielding an inversion problem that is tractable but informationally limited. 
Fundamentally, \textbf{No algorithm can recover what was already lost optically}. 

\subsubsection{Hessian matrix Analysis}
For a linear least-squares problem with objective function $f(\mathbf{I})$, the Hessian matrix $\nabla^2 f(\mathbf{I})$ is exactly the coefficient matrix $\mathbf{\Phi}^\top \mathbf{\Phi}$ from the normal equations. This matrix captures the correlations among the column vectors of the sensing matrix and dictates both the ill-posedness and convergence behavior of the optimization problem. As illustrated in Fig. \ref{Fig4}, a comparative analysis of the sensing matrices and Hessian matrices offers further insight into the inherent optimization characteristics of CSI systems based on different principles.

In the Hessian matrix $\mathbf{\Phi}^\top \mathbf{\Phi}$, nonzero off-diagonal elements indicate strong coupling among parameters. This coupling leads to a broad eigenvalue distribution and a high condition number $\kappa(\mathbf{\Phi}^\top \mathbf{\Phi}) \gg 1$, rendering the optimization problem numerically ill-conditioned. As a result, the gradient direction deviates from the steepest descent path toward the optimum, and the convergence path exhibits oscillatory “zig-zag” behavior, requiring numerous iterations to reconcile parameter interactions. Furthermore, explicitly forming $\mathbf{\Phi}^\top \mathbf{\Phi}$ incurs a computational complexity of $\mathcal{O}(N^3)$ and a memory footprint of $\mathcal{O}(N^2)$. For typical SDI problems, $N$ can reach millions, making direct inversion computationally infeasible.


\subsubsection{Diagonalisation in Frequency Domain }


We observe that the off-diagonal structure of the Hessian matrix in SDI systems arises entirely from the convolutional encoding induced by the PSFs. 
A key insight is that this convolution is equivalent to a straightforward multiplication in the frequency domain via the optical transfer function (OTF).
This transform-domain representation serves as a foundational prior, enabling the design of efficient iterative frameworks for stable and rapid reconstruction. 
Returning to the SDI formulation, the relation $\mathbf{J} = \mathbf{\Phi}_1 \mathbf{I}$ in Eq.~\ref{3-1-3} describes the modulation of the input signal by spectrally variant PSFs. Here, $\mathbf{\Phi}_1$ is a block-circulant convolution matrix, which can be diagonalized by the Fourier transform:
\begin{equation}
\setlength{\abovedisplayskip}{0cm}
\setlength{\belowdisplayskip}{0cm}
    {{\bf{\Phi }}_{\bf{1}}} = {{\cal F}^{-1}}\Lambda {\cal F}
\label{3-12-5}
\end{equation}
\noindent
where $\mathcal{F}$ denotes the Fourier transform matrices, ${{\cal F}^{-1}}$ is the inverse discrete Fourier transform (IDFT) matrices. and $\Lambda$ is a diagonal matrix whose elements correspond to the Fourier transform of the PSF (i.e., the OTF).

Consequently, the Hessian matrix can be expressed in the frequency domain as:
\begin{equation}
\setlength{\abovedisplayskip}{0cm}
\setlength{\belowdisplayskip}{0cm}
    {\bf{\Phi }}_1^{\top}{{\bf{\Phi }}_{\bf{1}}} = {{\cal F}^{ - 1}}{\left| \Lambda  \right|^2}{\cal F}
\label{3-12-6}
\end{equation}
\noindent
where $\left| \Lambda  \right|^2$ is a diagonal matrix containing the power spectrum of the OTF. If we consider the subproblem $\mathop {\min }\limits_{\bf{I}} \frac{1}{2}{\left\| {{\bf{J}} - {{\bf{\Phi }}_{\bf{1}}}{\bf{I}}} \right\|^2}$, it can be transformed into the frequency domain as follows:
\begin{equation}
\setlength{\abovedisplayskip}{0cm}
\setlength{\belowdisplayskip}{0cm}
    \mathop {\min }\limits_{{{\bf{I}}^{\cal F}}} \frac{1}{2}{\left\| {{{\bf{J}}^{\cal F}} - \Lambda {{\bf{I}}^{\cal F}}} \right\|^2}
\label{3-12-7}
\end{equation}
\noindent
where ${{\bf{J}}^{\cal F}} = {\cal F}({\bf{J}})$, ${{\bf{I}}^{\cal F}} = {\cal F}({\bf{I}})$.
In contrast, inverting a diagonal Hessian matrix is computationally efficient: it requires only element-wise inversion of the diagonal entries, with a complexity of $\mathcal{O}(N)$. The solution to the optimization problem in the frequency domain is given by:
\begin{equation}
\setlength{\abovedisplayskip}{0cm}
\setlength{\belowdisplayskip}{0cm}
    {{{\bf{\hat I}}}^{\cal F}} = \frac{{{\Lambda ^{ - 1}}{{\bf{I}}^{\cal F}}}}{{{{\left| \Lambda  \right|}^2}}}
\label{3-12-8}
\end{equation}

Compared to the $\mathcal{O}(N^3)$ complexity of spatial-domain inversion, frequency-domain processing substantially reduces computational cost. Moreover, it significantly reduces the condition number of the Hessian matrix, leading to faster convergence and improved robustness to noise. Motivated by this insight, we decompose the SDI inverse problem into two complementary subproblems in different domains to exploit the diagonal structure of their respective Hessian matrices. This approach forms the basis of our efficient reconstruction framework: the Hierarchical Spatial-Frequency Aggregation Unfolding Framework (HSFAUF).

\begin{figure*}[t]
    \begin{center}
    \includegraphics[width=1\linewidth]{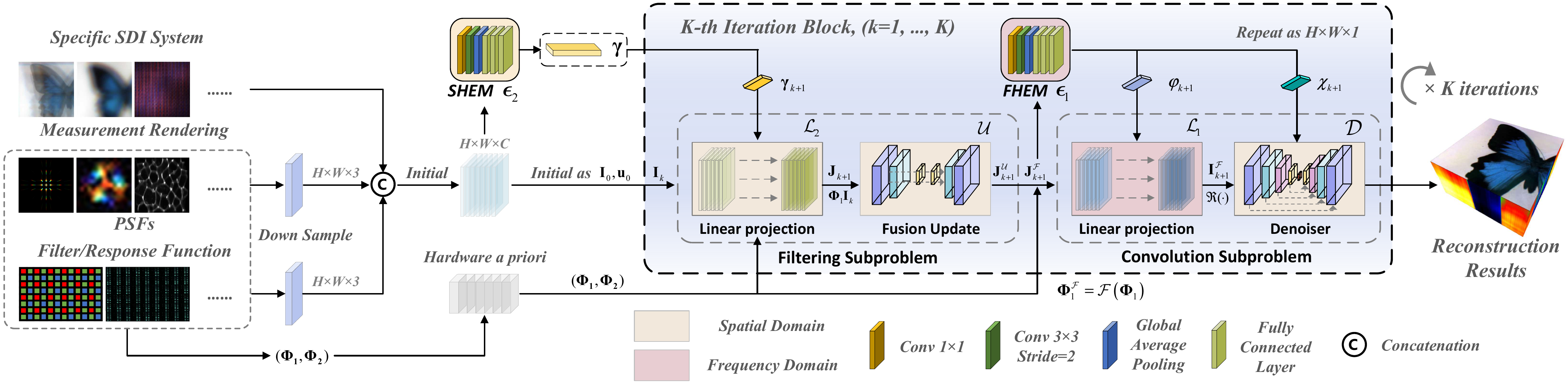}
    \end{center}
\vspace{-0.2cm} 
\caption{Illustration of HSFAUF architecture with k stages. Realize guided reconstruction by hierarchical extracting key cues from hardware a priori and imaging process characterization in the spatial and frequency domains.}
\label{Fig5}
\end{figure*}

\subsection{Hierachical Spatial-Frequency Aggregation Unfolding Framework, HSFAUF}
\label{subsec:HSFAUF}

SDI typically employs joint encoding via convolution kernels and filters, making it difficult to solve. To address the problem efficiently, we formulate a principled HSFAUF based on Maximum a Posteriori (MAP) theory as shown in Fig.~\ref{Fig5}. HSFAUF minimizes the energy function:
\begin{equation}
\setlength{\abovedisplayskip}{0cm}
\setlength{\belowdisplayskip}{0cm}
    {\bf{\hat I}} = \mathop {\min }\limits_{\bf{I}} \frac{1}{2}{\left\| {{\bf{M}} - {{\bf{\Phi }}_2}{{\bf{\Phi }}_1}{\bf{I}}} \right\|^{\bf{2}}} + {\bf{\lambda R}}({\bf{I}})
\label{eq3.2.1}
\end{equation}
\noindent
where ${\bf{R}}$ denotes a regularization term that may not necessarily be differentiable. By introduced auxiliary variable $\bf{J} = {\bf{\Phi _1}} \bf{I}$, Equation~\ref{eq3.2.1} is reformulated as a constrained optimization problem. To decouple the data fidelity and regularization terms:
\begin{equation}
\setlength{\abovedisplayskip}{0cm}
\setlength{\belowdisplayskip}{0cm}
    \langle \widehat {\bf{I}},\widehat {\bf{J}}\rangle  = \mathop {\min }\limits_{{\bf{I}},{\bf{J}}} \frac{1}{2}{\left\| {{\bf{M}} - {{\bf{\Phi }}_2}{\bf{J}}} \right\|^2} + \lambda {\bf{R}}({\bf{I}}),{\rm{s}}{\rm{.t}}{\rm{. }}{\bf{J}} = {{\bf{\Phi }}_1}{\bf{I}}
\label{eq3.2.2}
\end{equation}

Using half-quadratic splitting(HQS), Equation~\ref{eq3.2.2} is transformed into an unconstrained problem:
\begin{equation}
\setlength{\abovedisplayskip}{0cm}
\setlength{\belowdisplayskip}{0cm}
    \langle {\bf{\hat I}},{\bf{\hat J}}\rangle  \!=\! \mathop {\min }\limits_{{\bf{I}},{\bf{J}}} \frac{1}{2}{\left\| {{\bf{M}} \!-\! {{\bf{\Phi }}_2}{\bf{J}}} \right\|^2} \!+\! \lambda {\bf{R}}({\bf{I}}) \!+\! \frac{\gamma }{2}{\left\| {{\bf{J}} \!-\! {{\bf{\Phi }}_1}{\bf{I}}} \right\|^2}
\label{eq3.2.3}
\end{equation}
\noindent
where $\mu$ is a penalty parameter. Thus, Equation~\ref{eq3.2.3} can be  split into two subproblems for iterative solving. 
\begin{equation}
\setlength{\abovedisplayskip}{0cm}
\setlength{\belowdisplayskip}{0cm}
    {{\bf{J}}_{k + 1}} = \mathop {\arg \min }\limits_{\bf{J}} \frac{1}{2}{\left\| {{\bf{M}} - {{\bf{\Phi }}_2}{\bf{J}}} \right\|^2} + \frac{\gamma }{2}{\left\| {{\bf{J}} - {{\bf{\Phi }}_1}{{\bf{I}}_k}} \right\|^2}
\label{eq3.2.4}
\end{equation}

\begin{equation}
    {{\bf{I}}_{k + 1}} = \mathop {\arg \min }\limits_{\bf{I}} \frac{{\bf{\gamma }}}{2}{\left\| {{{\bf{J}}_{k + 1}} - {{\bf{\Phi }}_1}{\bf{I}}} \right\|^2} + {\bf{\lambda R}}({\bf{I}})
\label{eq3.2.5}
\end{equation}
\noindent
where ${{{\bf{J}}_k}}$ and ${{{\bf{I}}_k}}$ represent the $k-th$ HQS iteration respectively. 
According to Equation \ref{3-12-1}, we can diagonalise ${\bf{\Phi }}_1$ in the frequency domain.
Let ${\bf{\Phi }}_1^\mathcal{F}$ represent ${\cal F}\left( {{{\bf{\Phi }}_1}} \right)$, leading to ${\bf{\Phi }}_1^\mathcal{F}\mathop  = \limits^{{\rm{def}}} diag\{ {\psi _1}, \cdots ,{\psi _{nc}}\}  = {\bf{\Phi }}_1^{{\mathcal{F}^\top}}$. To fully exploit the spatial-frequency properties of the SDI inverse problem, we solve Equation~\ref{eq3.2.4} and Equation~\ref{eq3.2.5} as the filtering subproblem and the convolution subproblem respectively.

\subsubsection{Filtering Subproblem} 

For Equation~\ref{eq3.2.4}, a closed-form solution can be derived straightforwardly by setting the gradient with respect to $\mathbf{J}$ to zero:

\begin{equation}
\setlength{\abovedisplayskip}{0cm}
\setlength{\belowdisplayskip}{0cm}
    {{\bf{J}}_{k + 1}} = {({\bf{\Phi }}_2^\top{{\bf{\Phi }}_2} + \gamma \bf{1})^{ - 1}}({\bf{\Phi }}_2^\top{\bf{M}} + \gamma {{\bf{\Phi }}_1}{{\bf{I}}_k})
\label{eq3.2.6}
\end{equation}
\noindent
where $\bf{1}$ is an identity matrix. Here, we can further reduce the computational complexity of ${({\bf{\Phi }}_2^\top{{\bf{\Phi }}_2} + \gamma \bf{1})^{ - 1}}$ by employing the matrix inverse formula:

\begin{equation}
\setlength{\abovedisplayskip}{0cm}
\setlength{\belowdisplayskip}{0cm}
    {({\bf{\Phi }}_2^\top{{\bf{\Phi }}_2} \!+\! \gamma {\bf{1}})^{ - 1}} \!=\! {\gamma ^{ - 1}}{\bf{1}} \!-\! {\gamma ^{ - 1}}{\bf{\Phi }}_2^\top{({\bf{1}} \!+\! {{\bf{\Phi }}_2}{\gamma ^{ - 1}}{\bf{\Phi }}_2^\top)^{ - 1}}{{\bf{\Phi }}_2}{\gamma ^{ - 1}}
\label{eq3.2.7}
\end{equation}

By plugging Equation~\ref{eq3.2.7} into Equation~\ref{eq3.2.6}. (6), we can reformulate Equation~\ref{eq3.2.6} as:

\begin{equation}
\setlength{\abovedisplayskip}{0cm}
\setlength{\belowdisplayskip}{0cm}
    \left\{ {\begin{array}{*{20}{c}}
    {{{\bf{J}}_{k + 1}} = \frac{{{\bf{\Phi }}_2^\top{\bf{M}} + \gamma {{\bf{\Phi }}_1}{{\bf{I}}_k}}}{\gamma } - \frac{{{\bf{\Phi }}_2^\top{\cal B}{{\bf{\Phi }}_2}{{\bf{\Phi }}_1}{{\bf{I}}_k}}}{\gamma } - \frac{{{\bf{\Phi }}_2^\top{\cal C}{\bf{M}}}}{{{\gamma ^2}}}}\\
    {{\cal B} = ({\bf{1}} + {{\bf{\Phi }}_2}{\gamma ^{ - 1}}{\bf{\Phi }}_2^\top)}^{-1}\\
    {{\cal C} = {\bf{\Phi }}_2^\top{{({\bf{1}} + {{\bf{\Phi }}_2}{\gamma ^{ - 1}}{\bf{\Phi }}_2^\top)}^{ - 1}}{{\bf{\Phi }}_2}{\bf{\Phi }}_2^\top}
    \end{array}} \right.
\label{eq3.2.8}
\end{equation}
\noindent
where ${{\bf{\Phi }}_2}$ represents a linear process encompassing pixel-wise filtering and band-wise integration. As a result, the Hessian matrix ${{\bf{\Phi }}_2}{\bf{\Phi }}_2^\top \in {{\rm{R}}^{n \times n}}$ for the filtering subproblem is strictly diagonal. Let ${{\bf{\Phi }}_2}{\bf{\Phi }}_2^\top = diag\{ {\eta _1},...,{\eta _n}\}$, we then obtain:

\begin{equation}
\setlength{\abovedisplayskip}{0cm}
\setlength{\belowdisplayskip}{0cm}
    \begin{array}{*{20}{c}}
    {{\cal B} = diag\{ \frac{\gamma }{{\gamma  + {\eta _1}}},...,\frac{\gamma }{{\gamma  + {\eta _n}}}\} }\\
    {{\cal C} = diag\{ \frac{{\gamma {\eta _1}}}{{\gamma  + {\eta _1}}},...,\frac{{\gamma {\eta _n}}}{{\gamma  + {\eta _n}}}\} }
    \end{array}
\label{eq3.2.9}
\end{equation}

The solution can thus be obtained in an element-wise manner. Notably, in solving the filtering subproblem, we treat ${{\bf{\Phi }}_1}{{\bf{I}}_k}$ as a fixed vector while focusing primarily on the structure of ${{\bf{\Phi }}_2}$. Let ${\bf{M}} \mathop  = \limits^{def} [M_1, \dots, M_n]^\top$ and denote the $i$-th element of ${{\bf{\Phi }}_2}{{\bf{\Phi }}_1}{{\bf{I}}_k}$ as $[{{\bf{\Phi }}_2}{{\bf{\Phi }}_1}{{\bf{I}}_k}]_i$. We plug Equation~\ref{eq3.2.9} into Equation~\ref{eq3.2.8} as:

\begin{equation}
\setlength{\abovedisplayskip}{0cm}
\setlength{\belowdisplayskip}{0cm}
    {{\bf{J}}_{k + 1}} \!=\! {{\bf{\Phi }}_1}{{\bf{I}}_k} \!+\! {\bf{\Phi }}_2^\top{[\frac{{{M_1} \!-\! {{[{{\bf{\Phi }}_2}{{\bf{\Phi }}_1}{{\bf{I}}_k}]}_1}}}{{\gamma  \!+\! {\eta _1}}},...,\frac{{{M_n} \!-\! {{[{{\bf{\Phi }}_2}{{\bf{\Phi }}_1}{{\bf{I}}_k}]}_n}}}{{\gamma  \!+\! {\eta _n}}}]^\top}
\label{eq3.2.10}
\end{equation}
\noindent
where $\left\{ {{M_i} - {{[{{\bf{\Phi }}_2}{{\bf{\Phi }}_1}{{\bf{I}}_k}]}_i}} \right\}_{i = 1}^n$ can be efficiently updated by ${\bf{M}} - {{\bf{\Phi }}_2}{{\bf{\Phi }}_1}{{\bf{I}}_k}$.

It is noteworthy that as the number of iterations increases, ${{\bf{J}}_{k + 1}}$ and ${{\bf{\Phi }}_1}{{\bf{I}}_k}$ are expected to converge. However, in practice, they rarely achieve exact agreement. A practical solution is to introduce a \textbf{F}usion \textbf{U}pdate \textbf{M}odule (FUM) that aggregates both signals to improve stability:

\begin{equation}
\setlength{\abovedisplayskip}{0cm}
\setlength{\belowdisplayskip}{0cm}
    {\bf{J}}_{k + 1}^{\cal U} = {\cal U}({{\bf{J}}_{k + 1}},{{\bf{\Phi }}_1}{{\bf{I}}_k})
\label{eq3.2.11}
\end{equation}
\noindent

Building upon the foregoing analysis, set ${\boldsymbol{\gamma}}\stackrel{\text{def}}{=}[\gamma_{1},\cdots,\gamma_{k}]
$,  the filter subproblem is formulated as:

\begin{equation}
\setlength{\abovedisplayskip}{0cm}
\setlength{\belowdisplayskip}{0cm}
    \left\{ {\begin{array}{*{20}{c}}
    {{\boldsymbol{\gamma }} = \left( {{\bf{M}},{{\bf{\Phi }}_1},{{\bf{\Phi }}_2}} \right),}\\
    {{{\bf{J}}_{k + 1}} = {{\cal L}_2}\left( {{\bf{M}},{{\bf{\Phi }}_1},{{\bf{\Phi }}_2},{{\bf{\gamma}}_{k + 1}}}, {{\bf{I}}_k} \right),}\\
    {{\bf{J}}_{k + 1}^{\cal U} = {\cal U}({{\bf{J}}_{k + 1}},{{\bf{\Phi }}_1}{{\bf{I}}_k});}
    \end{array}} \right.
\label{eq3.2.11}
\end{equation}
\noindent

\noindent
where ${\boldsymbol{\epsilon}}_{2}$ represents a parameter estimator, ${\mathcal{L}}_{2}$ corresponds to the linear projection in Equation~\ref{eq3.2.10}.

\subsubsection{Convolution Subproblem}

Since ${\cal F}\left( {\bf{J}} \right) = {\cal F}({{\bf{\Phi }}_1}){\cal F}({\bf{I}})$, that is ${{\bf{J}}^{\cal F}} = {\bf{\Phi }}_1^{\cal F}{{\bf{I}}^{\cal F}}$, ${\bf{J}}^{\mathcal{F}}$ and ${\bf{I}}^{\mathcal{F}}$ are the frequency domain representations of $\bf{J}$ and $\bf{I}$ respectively, thus ${\bf{\Phi }}_1^{\cal F}$ represents the element-wise multiplication in the frequency domain, and ${\bf{\Phi }}_1^{\cal F}\mathop  = \limits^{{\rm{def}}} diag\{ {\psi _1}, \cdots ,{\psi _{nc}}\}  = {\bf{\Phi }}_1^{{{\cal F}^\top}} $. Transforming Equation~\ref{eq3.2.5} to the frequency domain gives:
\begin{equation}
\setlength{\abovedisplayskip}{0.1cm}
\setlength{\belowdisplayskip}{0.1cm}
    {\bf{I}}_{k + 1}^{\cal F} = \mathop {\arg \min }\limits_{{{\bf{I}}^{\cal F}}} \frac{\gamma }{2}{\left\| {{\bf{J}}_{k + 1}^{\cal F} - {\bf{\Phi }}_1^{\cal F}{{\bf{I}}^{\cal F}}} \right\|^2} + \lambda {{\bf{R}}_1}\left( {{{\bf{I}}^{\cal F}}} \right)
\label{eq.3.2.12}
\end{equation}

\noindent
where we set ${\bf{J}}_{k+1}^{\mathcal{F}} = \mathcal{F}\left({\bf{J}}_{k + 1}^{\cal U}\right)$ to ensure iterative continuity.
For Equation~\ref{eq.3.2.12}, we introduce auxiliary variable $\bf{u}\in\mathbb{R}^{nC\times1}$ subject to $\bf{I}\in\mathbb{R}^{nC\times1}$ to decouple the convolution subproblem and apply the HQS method, leading to the unconstrained problem:
\begin{equation}
\footnotesize{
    \left\langle {{{{\bf{\hat I}}}^{\cal F}},{{{\bf{\hat u}}}^{\cal F}}} \right\rangle  \!=\! \mathop {\min }\limits_{{{\bf{I}}^{\cal F}},{{\bf{u}}^{\cal F}}} \frac{\gamma }{2}{\left\| {{\bf{\Phi }}_1^{\cal F}{{\bf{I}}^{\cal F}} \!-\! {\bf{J}}_{k + 1}^{\cal F}} \right\|^2} \!+\! \frac{\mu }{2}{\left\| {{{\bf{I}}^{\cal F}} \!-\! {{\bf{u}}^{\cal F}}} \right\|^2} \!+\! \lambda {{\bf{R}}_1}({{\bf{u}}^{\cal F}})
}
\label{eq.3.2.13}
\end{equation}
\noindent
where the ${{\bf{u}}^{\cal F}}$ is the frequency domain representation of ${\bf{u}}$, and the consistency between $\mathbf{I}^{\mathcal{F}}$, and $\mathbf{u}$ is enforced via parameter $\mu$. The convolution subproblem decomposes into:

\begin{equation}
\setlength{\abovedisplayskip}{0cm}
\setlength{\belowdisplayskip}{0cm}
    {\bf{I}}_{k + 1}^{\cal F} \!=\! \mathop {\arg \min }\limits_{{{\bf{I}}^{\cal F}}} \frac{\gamma }{2}{\left\| {{\bf{\Phi }}_1^{\cal F}{{\bf{I}}^{\cal F}} \!-\! {\bf{J}}_{k + 1}^{\cal F}} \right\|^2} \!+\! \frac{\mu }{2}{\left\| {{{\bf{I}}^{\cal F}} \!-\! {\bf{u}}_k^{\cal F}} \right\|^2}
\label{eq.3.2.14}
\end{equation}

\begin{equation}
\setlength{\abovedisplayskip}{0cm}
\setlength{\belowdisplayskip}{0cm}
    {\bf{u}}_{k + 1}^{\cal F} = \mathop {\arg \min }\limits_{{{\bf{u}}^{\cal F}}} \frac{\mu }{2}{\left\| {{\bf{I}}_{k + 1}^{\cal F} - {{\bf{u}}^{\cal F}}} \right\|^2} + \lambda {{\bf{R}}_1}({{\bf{u}}^{\cal F}})
\label{eq.3.2.15}
\end{equation}
\noindent
where the data fidelity term relates to a least-squares problem, so $\mathbf{I}_{k+1}^{\mathcal{F}}$ in equation~\ref{eq.3.2.14} has a closed-form solution:

\begin{align}
    \mathbf{I}_{k+1}^{\cal F} &= {\left( {\gamma {\bf{\Phi }}{{_1^{\cal F}}^\top}{\bf{\Phi }}_1^{\cal F} + \mu {\bf{1}}} \right)^{ - 1}}\left( {\gamma {\bf{\Phi }}{{_1^{\cal F}}^\top}{\bf{J}}_{k + 1}^{\cal F} + \mu {\bf{u}}_k^{\cal F}} \right) \notag \\
    & = {\left( {{\bf{\Phi }}_1^{{{\cal F}}}{\bf{\Phi }}_1^{\cal F} \!+\! \varphi {\bf{1}}} \right)^{ - 1}}\left( {{\bf{\Phi }}_1^{{{\cal F}}}{\bf{J}}_{k + 1}^{\cal F} \!+\! \varphi {\bf{u}}_k^{\cal F}} \right)
\label{eq.3.2.16}
\end{align}

\noindent
where $\varphi = \frac{\mu}{\gamma}$. 
Given ${\bf{\Phi }}_1^{\cal F}\mathop  = \limits^{{\rm{def}}} diag\{ {\psi _1}, \cdots ,{\psi _{nc}}\}$, the closed-form solution in the frequency domain, as expressed in Equation~\ref{eq.3.2.16}, can be efficiently computed in an element-wise manner:

\begin{equation}
\setlength{\abovedisplayskip}{0cm}
\setlength{\belowdisplayskip}{0cm}
    {\bf{I}}_{k + 1}^{\cal F} = {[\frac{{{{[{\bf{\Phi }}_1^{\cal F}{\bf{J}}_{k + 1}^{\cal F} \!+\! \varphi {\bf{u}}_k^{\cal F}]}_1}}}{{\varphi  \!+\! \psi _1^2}},...,\frac{{{{[{\bf{\Phi }}_1^{\cal F}{\bf{J}}_{k + 1}^{\cal F} \!+\! \varphi {\bf{u}}_k^{\cal F}]}_{nC}}}}{{\varphi  \!+\! \psi _{nC}^2}}]^\top}
\label{eq.3.2.17}
\end{equation}
\noindent
where $\left. {\{ {{[{\bf{\Phi }}_1^{\cal F}{\bf{J}}_{k + 1}^{\cal F} \!+\! \varphi {\bf{u}}_k^{\cal F}]}_i}\} } \right|_{i = 1}^{nC}$ can be efficiently updated by ${{\bf{\Phi }}_1^{\cal F}{\bf{J}}_{k + 1}^{\cal F} + \varphi {\bf{u}}_k^{\cal F}}$.

From a Bayesian perspective, Equation~\ref{eq.3.2.15} is equivalent to denoising an image corrupted by Gaussian noise at noise level $\sqrt {{\lambda _{k + 1}}/{\mu _{k + 1}}}$ . However, because frequency-domain denoising requires greater computational and storage resources, while its fundamental goal mirrors that of spatial-domain denoising. We can remap the Equation~\ref{eq.3.2.15} back to the spatial domain to improve efficiency. It should be noted, however, that the complex division in Equation~\ref{eq.3.2.17} can disrupt Hermitian symmetry, which is necessary to ensure real-valued image representations. To avoid the numerical issues caused by this symmetry breaking, we compute $\mathbf{I}_{k+1}$ as follow:

\begin{equation}
\setlength{\abovedisplayskip}{0cm}
\setlength{\belowdisplayskip}{0cm}
    {{\bf{I}}_{k + 1}} = \Re ({{\cal F}^{ - 1}}({\bf{I}}_{k + 1}^{\cal F}))
\label{eq.3.3.1}
\end{equation}

\noindent
where the operator $\Re ( \cdot )$ denotes the real part of a complex-valued variable.

So Equation~\ref{eq.3.2.15} can be reformulated in the spatial domain:

\begin{equation}
\setlength{\abovedisplayskip}{0cm}
\setlength{\belowdisplayskip}{0cm}
    {{\bf{u}}_{k + 1}} = \mathop {\arg \min }\limits_{\bf{u}} \frac{\mu }{2}{\left\| {{\bf{I}}{}_{k + 1} - {\bf{u}}} \right\|^2} + \xi {{\bf{R}}_1}({\bf{u}})
\label{eq.3.2.18.3}
\end{equation}

We also set $\xi$ as iteration-specific parameters and Equation~\ref{eq.3.2.18.3} can be reformuted as:
\begin{equation}
\setlength{\abovedisplayskip}{0cm}
\setlength{\belowdisplayskip}{0cm}
    {{\bf{u}}_{k + 1}} = \mathop {\arg \min }\limits_{\bf{u}} \frac{1}{{2{{\left( {\sqrt {\frac{{{{\bf{\xi }}_{k + 1}}}}{{{{\bf{\mu }}_{k + 1}}}}} } \right)}^2}}}{\left\| {{{\bf{I}}_{k + 1}} - {\bf{u}}} \right\|^2} + {{\bf{R}}_1}({\bf{u}})
\label{eq.3.2.19}
\end{equation}
\noindent
thus it can be interpreted as a Gaussian denoising problem with noise level $\sqrt{\frac{\mathbf{\xi}_{k+1}}{\mathbf{\mu}_{k+1}}}$. Similarly, $\frac{\mathbf{\mu}_{k+1}}{\mathbf{\xi}_{k+1}}$ acts as a parameter estimator derived from SDI. Set $\boldsymbol{\varphi}\stackrel{\text{def}}{=}\left[{\varphi}_{1},\cdots,\varphi_{k}\right]$, $\chi_{k} = \frac{\mu_{k}}{\xi_{k}}$, $\boldsymbol{\chi}\stackrel{\text{def}}{=} \left[\chi_{1},\cdots,\chi_{n}\right]$, so the convolution subproblem is:
\begin{equation}
\left\{
\begin{array}{l}
\setlength{\abovedisplayskip}{-0.1cm}
\setlength{\belowdisplayskip}{-0.1cm}
    ({\varphi _{k + 1}},{\chi _{k + 1}}) = \boldsymbol{\epsilon}_{1}\left( {{\bf{J}}_{k + 1}^{\cal F},{\bf{\Phi }}_1^{\cal F}} \right)\\
    \mathbf{I}_{k+1}^{\mathcal{F}} = \mathcal{L}_{1}\left( {{\bf{J}}_{k + 1}^{\cal F},{\bf{u}}_k^{\cal F},{\bf{\Phi }}_1^{\cal F},{\varphi _{k + 1}}} \right)\\
    {{\bf{I}}_{k + 1}} = \Re ({{\cal F}^{ - 1}}({\bf{I}}_{k + 1}^{\cal F}))\\
    \mathbf{u}_{k+1} = {{\cal D}}\left( {{{\bf{I}}_{k + 1}},{\chi _{k + 1}}} \right)
\end{array}  
\right.
\label{eq.3.2.20}
\end{equation}
\noindent
where $\boldsymbol{\epsilon}_{1}$ represents the parameter estimator computed from ${\bf{J}}_{k+1}^{\mathcal{F}}$ and ${{\bf{\Phi }}_1^{\cal F}}$. $\mathcal{L}_{1}$ represents the linear update in Equation~\ref{eq.3.2.17}. $\mathcal{D}$ represents the Gaussian denoising relate to Equation~\ref{eq.3.2.19}.

In summary, the HSFAUF enables efficient SDI inverse problem solving by alternately solving the filtering and convolutional subproblems. Notably, even when an SDI system uses a monochromatic camera, a spatially consistent filtering matrix can be constructed based on the camera’s quantum efficiency, making HSFAUF applicable to a wide range of SDI configurations. To facilitate understanding of the structure and iterative workflow of HSFAUF, we provide its pseudocode in Algorithm~\ref{algo1}.

\begin{figure*}[t]
    \begin{center}
    \includegraphics[width=1\linewidth]{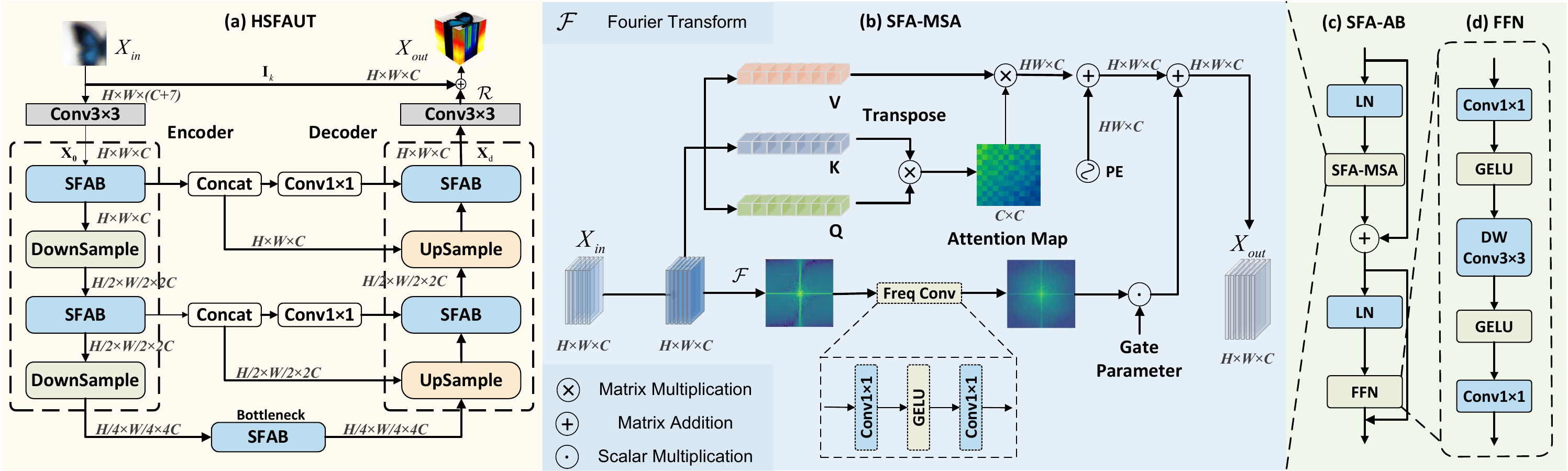}
    \end{center}
\vspace{-0.2cm} 
\caption{Illustration of the SFAT. Realize guided reconstruction by hierarchical extracting key cues from hardware a priori and imaging process characterization in the spatial and frequency domains.}
\vspace{-0.2cm} 
\label{algorithm}
\end{figure*}

\begin{algorithm}[h]
\caption{HSFAUF}\label{algo1}
\begin{algorithmic}[1]
\Require Measurement ${\bf{M}}$, sensing matrices $\mathbf{\Phi}_1, \mathbf{\Phi}_2$
\Ensure  Reconstructed HSIs singal ${{\bf{u}}_{k + 1}}$
\State \textbf{Initialization:}
\State Set ${{\bf{I}}_0} = {{\bf{u}}_0} = Initialize({\bf{M}},{{\bf{\Phi }}_1},{{\bf{\Phi }}_2})$
\State Calculate $\boldsymbol{\gamma} = {\boldsymbol{\epsilon}}_{2}\left( {{\bf{M}},{{\bf{\Phi }}_1},{{\bf{\Phi }}_2}} \right)$

\For{$k = 1$ to $K$}
    \State Solving Filtering Subproblem:
    \State ${{{\bf{J}}_{k + 1}} = {{\mathcal L}_2}\left( {{\bf{M}},{{\bf{\Phi }}_1},{{\bf{\Phi }}_2},{{\bf{\gamma}}_{k + 1}}}, {{\bf{I}}_k} \right)}$
    \State ${{\bf{J}}_{k + 1}^{\mathcal U} = {\cal U}({{\bf{J}}_{k + 1}},{{\bf{\Phi }}_1}{{\bf{I}}_k})}$
    \State Set ${\bf{\Phi }}_1^{\mathcal F} = {\cal F}\left( {{{\bf{\Phi }}_1}} \right)$, ${\bf{J}}_{k + 1}^{\cal F} = {\cal F}\left( {\bf{J}}_{k + 1}^{\cal U} \right)$
    \State Solving Convolution Subproblem:
    \State $({\varphi _{k + 1}},{\chi _{k + 1}}) = \boldsymbol{\epsilon}_{1}\left( {{\bf{J}}_{k + 1}^{\cal F},{\bf{\Phi }}_1^{\cal F}} \right)$
    \State $\mathbf{I}_{k+1}^{\mathcal{F}} = \mathcal{L}_{1}\left( {\bf{J}}_{k + 1}^{\cal F},{\bf{u}}_k^{\cal F},{\bf{\Phi }}_1^{\cal F},{\varphi _{k + 1}} \right)$ 
    \State Calculate ${{\bf{I}}_{k + 1}} = \Re ({{\cal F}^{ - 1}}({\bf{I}}_{k + 1}^{\mathcal F}))$
    \State $\mathbf{u}_{k+1} = {{\mathcal D}}\left( {{{\bf{I}}_{k + 1}},{\chi _{k + 1}}} \right)$ 
\EndFor

\State \Return $\mathbf{u}_{k+1}$ 

\State \textbf{Output:} Reconstructed signal $\mathbf{u}_{k+1}$ 
\end{algorithmic}
\end{algorithm}

\subsection{Spatial-Frequency Aggregation Transformer} 
Given that the SDI forward model couples spatial (filter function) and frequency-domain (OTF) encodings via Hadamard products, it is imperative that denoisers in a deep unfolding framework capture these cross-domain interactions.
Existing denoiser architectures, however, fail to provide efficient mechanisms for this purpose, primarily due to limited receptive fields and high computational complexity.

To address these challenges, we propose Spatial-Frequency Aggregation Transformer (SFAT) to play the role of denoisers. As shown in Fig.~\ref{algorithm}, SFAT adopts a three-level U-shaped structure built by the basic unit Spatial-Frequency Aggregation Attention Block (SFA-AB).

\textbf{SFAT Architecture:}
\textbf{Firstly}, SFAT uses a $conv 3 \times 3$ to map feature ${\bf{X}_{in}} \in {{\mathbb R}^{H \times W \times (C+7)}}$ that consists of ${\bf I}_k\in {{\mathbb R}^{H \times W \times C}}$ concatenated with stretched ${\chi _{k}} \in {{\mathbb R}^{H \times W \times 1}}$ and compressed OTFs feature into feature ${\bf{X}_0} \in {{\mathbb R}^{H \times W \times C}}$.
\textbf{Secondly}, ${{\bf X}_0}$ passes through the encoder, bottleneck, and decoder. Each level of the encoder or decoder contains an SFA-AB and a resizing module.
In Fig.~\ref{algorithm}(c), SFA-AB consists of two layer normalization (LN), a SFA-MSA, and a Feed-Forward Network (FFN) that is detailed in Fig.~\ref{algorithm}(d). The downsampling and upsampling modules are strided $conv4 \times 4$ and $deconv2 \times 2$. 
\textbf{Finally}, a $conv 3 \times 3$ operates on ${\bf{X}_d}$ to generate a residual image ${\cal R} \in {{\mathbb R}^{H \times W \times C}}$. The output denoised image ${\bf X}_{out}$ (i.e. ${\bf u}_k$) is obtained by the sum of ${{\bf I}_k}$ and reshaped $\mathcal R$.

\textbf{Spatial-Frequency Aggregation Multi-head Self Attention:}
The core element of SFA-AB is the proposed Spatial-Frequency Aggregation Multi-head Self-Attention (SFA-MSA) module. Fig.~\ref{algorithm}(b) depicts the structure of SFA-MSA used in SFAT. The input tokens of SFA-MSA are denoted as ${\bf{x}_{in}} \in {{\mathbb R}^{H \times W \times C}}$. 
Subsequently, 
${{\bf x}_{in}}$ is processed through two parallel branches for domain-specific feature extraction: a spatial-spectral branch and a frequency-spectral branch. The resulting representations are then aggregated via a gated summation mechanism, adaptively weighted by learnable parameters, to achieve effective cross-domain integration.

\textbf{In Spatial-Spectral Branch}, ${{\bf x}_{in}}$ is reshaped into tokens ${\bf x} \in {{\mathbb R}^{H W \times C}}$, Then ${\bf x}$ is linearly projected into query ${\bf Q} \in {{\mathbb R}^{H W \times C}}$ , key ${\bf K} \in {{\mathbb R}^{H W \times C}}$ , and value ${\bf V} \in {{\mathbb R}^{H W \times C}}$ as:
\begin{equation}
\setlength{\abovedisplayskip}{0cm}
\setlength{\belowdisplayskip}{0cm}
    {\bf{Q}} = {{\bf{x}}}{{\bf{W}}^{\bf{Q}}},{\bf K} = {{\bf{x}}}{{\bf{W}}^{\bf K}},{\bf V} = {{\bf{x}}}{{\bf{W}}^{\bf V}}
\label{eq.3.4.1}
\end{equation}
\noindent
where ${{\bf{W}}^{\bf{Q}}},{{\bf{W}}^{\bf{K}}},{{\bf{W}}^{\bf{V}}} \in {{\mathbb R}^{C \times C}}$ are learnable parameters.

In terms of the attention mechanism, the spatial–spectral branch adopts a tokenization scheme similar to MST~\cite{MST}, treating each spectral representation as a token and computing self-attention over each ${\bf H}_j$:
\begin{equation}
\setlength{\abovedisplayskip}{0cm}
\setlength{\belowdisplayskip}{0cm}
    {{\bf H}_j} = {{\bf{V}}_j}(softmax ({\alpha _j}{\bf{K}}_j^\top{{\bf{Q}}_j}))
\label{eq.3.4.2}
\end{equation}
\noindent
where the ${\alpha _j} \in{{\mathbb R}^1}$ is a learnable parameter to adapt the ${{\bf H}_j}$ by re-weighting the matrix multiplication. then the Spatial-Spectral Multi-head Self-Attention (SS-MSA) can be calculated with a position embedding:
\begin{equation}
\setlength{\abovedisplayskip}{0cm}
\setlength{\belowdisplayskip}{0cm}
    { \bf \rm{SS - MSA}}({\bf{x}}) = Concat({{\bf H}_j}){\bf{W}} + {f_{PE}}({\bf{V}})
\label{eq.3.4.3}
\end{equation}
\noindent
where ${\bf W} \in{{\mathbb R}^{C \times C}}$ are learnable parameters, $f_{PE}( \cdot )$ is the function to generate position embedding. It consists of two depth-wise conv3×3 layers, a GELU, and reshape operations.

\textbf{In Frequency-Spectral Branch}, ${{\bf x}_{in}}$ is first permuted as ${{\bf x}_{perm}} \in {{\mathbb R}^{C \times H \times W}}$ to align spectral dimensions. The two-dimensional real-valued fast Fourier transform (RFFT) under orthogonal normalization ${{\cal F}_{r2d}^{ortho}(\cdot)}$ is then applied  to obtain frequency components. The amplitude spectrum is extracted and processed by a dedicated frequency-domain convolution layer $f_\theta$. To handle dimensional mismatch caused by RFFT, bilinear interpolation is applied to restore spatial resolution. The resulting features are then permuted back to the original data format:
\begin{equation}
\setlength{\abovedisplayskip}{0cm}
\setlength{\belowdisplayskip}{0cm}
    {\bf \rm FS}({{\bf x}_{perm}}) \!=\! Perm[Interp({f_\theta }\left| {{\cal F}_{r2d}^{ortho}({{\bf x}_{perm}})} \right|)]
\label{eq.3.4.4}
\end{equation}

\noindent
where the $f_\theta$ consists of two $conv 1 \times 1$ and a GELU.

\textbf{Finally}, we get the output of SFA-MSA by: 
\begin{equation}
\setlength{\abovedisplayskip}{0cm}
\setlength{\belowdisplayskip}{0cm}
    {\rm SFA - MSA}({{\bf x}_{in}}) = {\rm SS - MSA}({\bf x}) + \beta  \cdot {\rm FS}({{\bf  x}_{perm}})
\label{eq.3.4.4}
\end{equation}

\noindent
where $\beta$ is a learnable gating parameter to scale the frequency-domain output. 
Through SFA-MSA, we efficiently aggregate informative cross-domain cues from both spatial and frequency representations, thereby enhancing reconstruction fidelity.

\begin{table*}[t]
\caption{Quantitative comparison of results from different methods on \textbf{amplitude-Coded} Systems, PSNR (dB) and SSIM are reported. Red and
blue colors distinguish the best, second-best results, respectively.}
\vspace{-0.2cm}
\begin{center}
\resizebox{180mm}{38mm}{
\begin{tabular}{c|ccccccccccccc}
\hline
\rowcolor[HTML]{EFEFEF} 
\textbf{Algorithm}                                                                   & \textbf{Params}                                 & \textbf{GFLOPS}                                                        & \textbf{S1}                           & \textbf{S2}                           & \textbf{S3}                           & \textbf{S4}                           & \cellcolor[HTML]{EFEFEF}\textbf{S5}   & \cellcolor[HTML]{EFEFEF}\textbf{S6}   & \cellcolor[HTML]{EFEFEF}\textbf{S7}   & \cellcolor[HTML]{EFEFEF}\textbf{S8}   & \cellcolor[HTML]{EFEFEF}\textbf{S9}   & \cellcolor[HTML]{EFEFEF}\textbf{S10}  & \cellcolor[HTML]{EFEFEF}\textbf{Avg}   \\ \hline
\rowcolor[HTML]{FFFFFF} 
\cellcolor[HTML]{EFEFEF}                                                             & \cellcolor[HTML]{FFFFFF}                        & \cellcolor[HTML]{FFFFFF}                                               & 32.70                                 & 29.61                                 & 25.59                                 & 32.95                                 & 28.35                                 & 28.89                                 & 27.36                                 & 32.05                                 & {\color[HTML]{333333} 35.79}          & 27.65                                 & \textbf{30.09}                         \\
\rowcolor[HTML]{FFFFFF} 
\multirow{-2}{*}{\cellcolor[HTML]{EFEFEF}\textbf{UNet}~\cite{Unet}}         & \multirow{-2}{*}{\cellcolor[HTML]{FFFFFF}23.32M} & \multirow{-2}{*}{\cellcolor[HTML]{FFFFFF}5.26}                        & {\color[HTML]{333333} 0.949}          & 0.900                                 & 0.874                                 & 0.832                                 & 0.892                                 & 0.879                                 & 0.853                                 & 0.910                                 & 0.927                                 & 0.896                                 & \textbf{0.891}                         \\ \hline
\rowcolor[HTML]{FFFFFF} 
\cellcolor[HTML]{EFEFEF}                                                            & \cellcolor[HTML]{FFFFFF}                        & \cellcolor[HTML]{FFFFFF}{\color[HTML]{333333} }                        & 28.11                                 & 27.67                                 & {\color[HTML]{333333} 23.05}          & 29.86                                 & 25.74                                 & 26.84                                 & 22.65                                 & 29.96                                 & {\color[HTML]{333333} 32.05}          & 24.99                                 & \textbf{27.09}                         \\
\rowcolor[HTML]{FFFFFF} 
\multirow{-2}{*}{\cellcolor[HTML]{EFEFEF}\textbf{MIRNet}~\cite{MIRNet}}     & \multirow{-2}{*}{\cellcolor[HTML]{FFFFFF}2.05M} & \multirow{-2}{*}{\cellcolor[HTML]{FFFFFF}{\color[HTML]{333333} 14.60}} & {\color[HTML]{333333} 0.876}          & 0.851                                 & 0.823                                 & 0.807                                 & 0.848                                 & 0.843                                 & 0.807                                 & 0.890                                 & 0.880                                 & 0.846                                 & \textbf{0.847}                         \\ \hline
\rowcolor[HTML]{FFFFFF} 
\cellcolor[HTML]{EFEFEF}                                                           & \cellcolor[HTML]{FFFFFF}                        & \cellcolor[HTML]{FFFFFF}{\color[HTML]{333333} }                        & 31.77                                 & {\color[HTML]{333333} 28.82}          & 25.32                                 & {\color[HTML]{333333} 32.36}          & 27.58                                 & 27.34                                 & 25.80                                 & 31.24                                 & 32.84                                 & 25.85                                 & \textbf{28.89}                         \\
\rowcolor[HTML]{FFFFFF} 
\multirow{-2}{*}{\cellcolor[HTML]{EFEFEF}\textbf{Lambda-Net}~\cite{l-net}}   & \multirow{-2}{*}{\cellcolor[HTML]{FFFFFF}32.73M}    & \multirow{-2}{*}{\cellcolor[HTML]{FFFFFF}{\color[HTML]{333333} 23.01}} & {\color[HTML]{333333} 0.925}          & 0.888                                 & 0.851                                 & 0.809                                 & 0.861                                 & 0.851                                 & 0.824                                 & 0.903                                 & 0.887                                 & 0.868                                 & \textbf{0.867}                         \\ \hline
\rowcolor[HTML]{FFFFFF} 
\cellcolor[HTML]{EFEFEF}                                                             & \cellcolor[HTML]{FFFFFF}                        & \cellcolor[HTML]{FFFFFF}{\color[HTML]{333333} }                        & 32.12                                 & 30.10                                 & {\color[HTML]{333333} 25.42}          & 32.23                                 & 27.75                                 & 28.20                                 & 25.48                                 & 32.68                                 & {\color[HTML]{333333} 36.05}          & 28.80                                 & \cellcolor[HTML]{FFFFFF}\textbf{29.89} \\
\rowcolor[HTML]{FFFFFF} 
\multirow{-2}{*}{\cellcolor[HTML]{EFEFEF}\textbf{MST++}~\cite{MST++}}       & \multirow{-2}{*}{\cellcolor[HTML]{FFFFFF}1.33M} & \multirow{-2}{*}{\cellcolor[HTML]{FFFFFF}{\color[HTML]{333333} 17.46}} & {\color[HTML]{333333} 0.943}          & 0.914                                 & 0.891                                 & 0.851                                 & 0.891                                 & 0.900                                 & 0.869                                 & 0.932                                 & 0.932                                 & 0.919                                 & \cellcolor[HTML]{FFFFFF}\textbf{0.904} \\ \hline
\rowcolor[HTML]{FFFFFF} 
\cellcolor[HTML]{EFEFEF}                                                            & \cellcolor[HTML]{FFFFFF}                        & \cellcolor[HTML]{FFFFFF}{\color[HTML]{333333} }                        & 32.94                                 & {\color[HTML]{333333} 29.89}          & 24.84                                 & 32.86                                 & 30.05                                 & 28.34                                 & 26.49                                 & 32.41                                 & {\color[HTML]{333333} 35.19}          & 28.15                                 & \textbf{30.12}                         \\
\rowcolor[HTML]{FFFFFF} 
\multirow{-2}{*}{\cellcolor[HTML]{EFEFEF}\textbf{TSA-Net}~\cite{TSA-net}}    & \multirow{-2}{*}{\cellcolor[HTML]{FFFFFF}44.25M}    & \multirow{-2}{*}{\cellcolor[HTML]{FFFFFF}{\color[HTML]{333333} 91.94}} & {\color[HTML]{333333} 0.950}          & 0.910                                 & 0.867                                 & 0.834                                 & 0.912                                 & 0.874                                 & 0.843                                 & 0.919                                 & 0.936                                 & 0.903                                 & \textbf{0.895}                         \\ \hline
\rowcolor[HTML]{FFFFFF} 
\cellcolor[HTML]{EFEFEF}                                                            & \cellcolor[HTML]{FFFFFF}                        & \cellcolor[HTML]{FFFFFF}{\color[HTML]{333333} }                        & 33.02                                 & 29.90                                 & {\color[HTML]{333333} 25.90}          & 32.54                                 & 29.09                                 & 28.20                                 & 24.65                                 & 32.63                                 & {\color[HTML]{333333} 35.57}          & 28.81                                 & \textbf{30.03}                         \\
\rowcolor[HTML]{FFFFFF} 
\multirow{-2}{*}{\cellcolor[HTML]{EFEFEF}\textbf{MPRNet}~\cite{MPRNet}}       & \multirow{-2}{*}{\cellcolor[HTML]{FFFFFF}2.96M} & \multirow{-2}{*}{\cellcolor[HTML]{FFFFFF}{\color[HTML]{333333} 77.64}} & {\color[HTML]{333333} 0.950}          & 0.906                                 & 0.898                                 & 0.866                                 & 0.910                                 & 0.898                                 & 0.862                                 & 0.933                                 & 0.939                                 & 0.921                                 & \textbf{0.908}                         \\ \hline
\rowcolor[HTML]{FFFFFF} 
\cellcolor[HTML]{EFEFEF}                                                            & \cellcolor[HTML]{FFFFFF}                        & \cellcolor[HTML]{FFFFFF}{\color[HTML]{333333} }                        & 34.66 & 31.36          & 27.21                                 & 33.71                                 & 30.33 & 29.73                                 & 27.14          & 34.05                                 & 37.62                                 & 30.06                                 & \textbf{31.59}                         \\
\rowcolor[HTML]{FFFFFF} 
\multirow{-2}{*}{\cellcolor[HTML]{EFEFEF}\textbf{Restormer}~\cite{restormer}}   & \multirow{-2}{*}{\cellcolor[HTML]{FFFFFF}15.12M}    & \multirow{-2}{*}{\cellcolor[HTML]{FFFFFF}{\color[HTML]{333333} 87.87}} & 0.968 & 0.938                                 & 0.922                                 & 0.885                                 & 0.923                                 & 0.925                                 & 0.883                                 & 0.948                                 & 0.948          & 0.948                                 & \textbf{0.929}                         \\ \hline
\rowcolor[HTML]{FFFFFF} 
\cellcolor[HTML]{EFEFEF}                                                             & \cellcolor[HTML]{FFFFFF}                        & \cellcolor[HTML]{FFFFFF}{\color[HTML]{333333} }                        & 33.80 & 33.20 & 28.41 & 34.13 & 30.58 & 31.07 & 27.48 & 34.77 & 37.82 & 31.82 & {\color[HTML]{000000} \textbf{32.31}}  \\
\rowcolor[HTML]{FFFFFF} 
\multirow{-2}{*}{\cellcolor[HTML]{EFEFEF}\textbf{CSST-9stg}~\cite{ADIS}}   & \multirow{-2}{*}{\cellcolor[HTML]{FFFFFF}6.56M}    & \multirow{-2}{*}{\cellcolor[HTML]{FFFFFF}{\color[HTML]{333333} 70.44}} & 0.965 & 0.961 & 0.942 & 0.891 & 0.931 & 0.940 & 0.892 & 0.954 & 0.955 & 0.958 & {\color[HTML]{000000} \textbf{0.939}}  \\ \hline


\rowcolor[HTML]{F1E9E1} 
\cellcolor[HTML]{F1E9E1}                                                             & \cellcolor[HTML]{F1E9E1}                        & \cellcolor[HTML]{F1E9E1}                                               & 34.72          & 34.81 & 30.38 & 34.17 & {\color[HTML]{000000} 31.42}          & 32.47 & 28.77 & 35.78 & {\color[HTML]{3531FF} \textbf{39.36}} & {\color[HTML]{3531FF} \textbf{34.11}} & {\color[HTML]{000000} \textbf{33.60}}  \\
\rowcolor[HTML]{F1E9E1} 
\multirow{-2}{*}{\cellcolor[HTML]{F1E9E1}\textbf{HSFAUT-3stg (Ours)}}    & \multirow{-2}{*}{\cellcolor[HTML]{F1E9E1}2.95M}    & \multirow{-2}{*}{\cellcolor[HTML]{F1E9E1}64.72}                             & 0.966                                 & 0.968 & 0.958 & 0.900 & 0.925 & 0.953 & 0.895 & 0.959 & {\color[HTML]{3531FF} \textbf{0.962}} & 0.967 & {\color[HTML]{000000} \textbf{0.945}}  \\ \hline


\rowcolor[HTML]{F1E9E1} 
\cellcolor[HTML]{F1E9E1}                                                            & \cellcolor[HTML]{F1E9E1}                        & \cellcolor[HTML]{F1E9E1}                                               & {\color[HTML]{3531FF} \textbf{35.15}}          & {\color[HTML]{3531FF} \textbf{35.01}} & {\color[HTML]{3531FF} \textbf{30.44}} & {\color[HTML]{3531FF} \textbf{35.30}} & {\color[HTML]{CB0000} \textbf{32.06}}          & {\color[HTML]{3531FF} \textbf{32.57}} & {\color[HTML]{CB0000} \textbf{29.32}} & {\color[HTML]{3531FF} \textbf{35.89}} & {\color[HTML]{CB0000} \textbf{39.64}} & {\color[HTML]{000000} 33.87} & {\color[HTML]{3531FF} \textbf{33.93}}  \\
\rowcolor[HTML]{F1E9E1} 
\multirow{-2}{*}{\cellcolor[HTML]{F1E9E1}\textbf{HSFAUT-4stg (Ours)}}    & \multirow{-2}{*}{\cellcolor[HTML]{F1E9E1}3.43M}    & \multirow{-2}{*}{\cellcolor[HTML]{F1E9E1}82.72}                             & {\color[HTML]{3531FF} \textbf{0.968}}                                 & {\color[HTML]{3531FF} \textbf{0.971}} & {\color[HTML]{3531FF} \textbf{0.959}} & {\color[HTML]{3531FF} \textbf{0.907}} & {\color[HTML]{3531FF} \textbf{0.934}} & {\color[HTML]{3531FF} \textbf{0.955}} & {\color[HTML]{3531FF} \textbf{0.899}} & {\color[HTML]{3531FF} \textbf{0.960}} & {\color[HTML]{CB0000} \textbf{0.964}} & {\color[HTML]{3531FF} \textbf{0.969}} & {\color[HTML]{3531FF} \textbf{0.949}}  \\ \hline


\rowcolor[HTML]{F1E9E1} 
\cellcolor[HTML]{F1E9E1}                                                           & \cellcolor[HTML]{F1E9E1}                        & \cellcolor[HTML]{F1E9E1}                                               & {\color[HTML]{CB0000}\textbf{36.02}}          & {\color[HTML]{CB0000} \textbf{35.78}} & {\color[HTML]{CB0000} \textbf{31.26}} & {\color[HTML]{CB0000} \textbf{35.44}} & {\color[HTML]{3531FF} \textbf{31.45}}          & {\color[HTML]{CB0000} \textbf{32.80}} & {\color[HTML]{3531FF} \textbf{29.01}} & {\color[HTML]{CB0000} \textbf{36.15}} & {\color[HTML]{000000} 38.55} & {\color[HTML]{CB0000} \textbf{34.96}} & {\color[HTML]{CB0000} \textbf{34.14}}  \\
\rowcolor[HTML]{F1E9E1} 
\multirow{-2}{*}{\cellcolor[HTML]{F1E9E1}\textbf{HSFAUT-5stg (Ours)}}   & \multirow{-2}{*}{\cellcolor[HTML]{F1E9E1}3.92M}    & \multirow{-2}{*}{\cellcolor[HTML]{F1E9E1}100.72}                             & {\color[HTML]{CB0000}\textbf{0.971}}                                 & {\color[HTML]{CB0000} \textbf{0.974}} & {\color[HTML]{CB0000} \textbf{0.968}} & {\color[HTML]{CB0000} \textbf{0.908}} & {\color[HTML]{CB0000} \textbf{0.936}} & {\color[HTML]{CB0000} \textbf{0.958}} & {\color[HTML]{CB0000} \textbf{0.911}} & {\color[HTML]{CB0000} \textbf{0.966}} & {\color[HTML]{000000} 0.958} & {\color[HTML]{CB0000} \textbf{0.973}} & {\color[HTML]{CB0000} \textbf{0.952}}  \\ \hline
\end{tabular}
}
\end{center}
\label{tab1}
\vspace{-0.2cm} 
\end{table*}


\subsection{Realization of HSFAUT} 
\vspace{-0.1cm}

We now detail the implementation of the HSFAUT. To incorporate imaging process priors from both spatial and frequency domains, we introduce two lightweight modules: the Spatial Hyperparameter Estimation Module (SHEM) and the Frequency Hyperparameter Estimation Module (FHEM). While sharing an identical architecture, they differ in their inputs and outputs. Specifically, SHEM estimates global bootstrap parameters to each iteration as $\boldsymbol{\epsilon}_{2}$, while FHEM learns the bootstrap parameters at each iteration step as $\boldsymbol{\epsilon}_{1}$, using ${{\bf{J}}_{k + 1}^{\cal F}}$ as input. 
Both share an identical lightweight architecture—comprising a 1×1 convolution, a 3×3 convolution, a global average pooling layer, and three fully-connected layers.
Notably, given the highly sparse nature of the PSFs $(512\times512\times28)$ and filtering functions $(256\times256\times28)$, when they are input to $\boldsymbol{\epsilon}_{2}$ as imaging physical priors, they are uniformly compressed to $256\times256\times3$ to enable more efficient parameter estimation.
Leveraging parameters from SHEM and FHEM, HSFAUF iteratively addresses the convolution and filtering subproblems, as detailed in Equations~\ref{eq3.2.11} and \ref{eq.3.2.20}, respectively. 

Additionally, the HSFAUF incorporates an adaptive Fusion Update Module (FUM), which progressively refines parameter updates through three cascaded SFA-AB blocks.


\section{Simulation Experiment}
\label{sec:method}

\subsection{Simulation Experiment Setup}
\label{subsec:Setup}



To demonstrate the superiority of HSFAUT in solving the inverse problems of all kinds of SDI systems, we compare HSFAUT with other state-of-the-art algorithms, Similar to previous works~\cite{TSA-net,MST,CST,ADIS,DAUHST}, we select 28 wavelengths ranging from 450$nm$ to 650$nm$ and derive them via spectral interpolation for HSIs.
In system selection, we have chosen three representative SDI configurations: \textbf{amplitude coding}~\cite{ADIS}, \textbf{phase coding}~\cite{DOE_Jeon}, and \textbf{scattering coding}~\cite{diffusercam}. Within the unified SDI reconstruction framework depicted in Figure~\ref{Fig3}, the PSFs for all three systems are standardized to a resolution of 512×512×28. Correspondingly, the filter functions are maintained at 256×256×28, aligning with the original HSI signals. A fair comparison is facilitated by substituting the PSFs and filter functions tailored for each specific system.

Similar to ADIS~\cite{ADIS}, we adopt two datasets, i.e., CAVE-1024~\cite{CAVE} and KAIST~\cite{KAIST} for simulation experiments. 10 scenes from the KAIST dataset are selected for testing, while the CAVE-1024 dataset and another 20 scenes from the KAIST dataset are selected for training. We implement all models by Pytorch. All HSFAUT models are trained with Adam~\cite{Adam} optimizer (${\beta _1}$ = 0.9 and ${\beta _2}$ = 0.999) using Cosine Annealing scheme~\cite{sgdr} for 300 epochs on an RTX 4090D GPU.

\begin{table*}[t]
\vspace{-0.1cm}
\caption{Quantitative comparison of reconstruction results from different algorithms on \textbf{amplitude-based SDI} systems, \textbf{phase-based SDI} systems and \textbf{scatter-Based SDI} systems, PSNR (dB), SSIM and SAM are reported. light blue shading denotes results achieved by non-HSFAUT methods, while beige shading indicates results from the proposed HSFAUT. For each SDI system, red, blue, and green colors distinguish the best, second-best, and third-best results, respectively. For detailed data about DOE-based system and diffuser-based system, please refer to the \textbf{supplementary materials}.}
\vspace{-0.3cm}
\begin{center}
\resizebox{2\columnwidth}{!}{
\begin{tabular}{c|ccc|ccc|ccc}
\hline
\rowcolor[HTML]{EFEFEF} 
\cellcolor[HTML]{EFEFEF}                  & \multicolumn{3}{c|}{\cellcolor[HTML]{EFEFEF}\textbf{ADIS}~\cite{ADIS}}        & \multicolumn{3}{c|}{\cellcolor[HTML]{EFEFEF}\textbf{DOE-Based}~\cite{DOE_Jeon}}                   & \multicolumn{3}{c}{\cellcolor[HTML]{EFEFEF}\textbf{Diffuser-Based}~\cite{diffusercam}}                                            \\ \cline{2-10} 
\rowcolor[HTML]{EFEFEF} 
\multirow{-2}{*}{\cellcolor[HTML]{EFEFEF}\textbf{Algorithm}} & \multicolumn{1}{l}{\cellcolor[HTML]{EFEFEF}PSNR/dB $\uparrow$} & \multicolumn{1}{l}{\cellcolor[HTML]{EFEFEF}SSIM $\uparrow$} & \multicolumn{1}{l|}{\cellcolor[HTML]{EFEFEF}SAM $\downarrow$} & \multicolumn{1}{l}{\cellcolor[HTML]{EFEFEF}PSNR/dB $\uparrow$} & \multicolumn{1}{l}{\cellcolor[HTML]{EFEFEF}SSIM $\uparrow$} & \multicolumn{1}{l|}{\cellcolor[HTML]{EFEFEF}SAM $\downarrow$} & \multicolumn{1}{l}{\cellcolor[HTML]{EFEFEF}PSNR/dB $\uparrow$} & \multicolumn{1}{l}{\cellcolor[HTML]{EFEFEF}SSIM $\uparrow$} & \multicolumn{1}{l}{\cellcolor[HTML]{EFEFEF}SAM $\downarrow$} \\ \hline
\rowcolor[HTML]{FFFFFF} 
\cellcolor[HTML]{EFEFEF}\textbf{UNet}~\cite{Unet}          & 30.09           & 0.891           & 6.789              & 32.88                    & 0.916       & 5.527          & 26.27             & 0.751       & 7.783                                         \\ \hline

\rowcolor[HTML]{FFFFFF} 
\cellcolor[HTML]{EFEFEF}\textbf{MIRNet}~\cite{MIRNet}       & 27.09 & 0.847 &8.195               & 29.91                                                   & 0.862     &6.500                                              & 20.59                                                   & {\color[HTML]{333333} 0.561}   & 10.971                      \\ \hline

\rowcolor[HTML]{FFFFFF} 
\cellcolor[HTML]{EFEFEF}\textbf{Lambda-Net}~\cite{l-net}    &28.89  &0.867  &7.032                & 30.28                                                   & 0.858     & 6.249                                             & {\color[HTML]{333333} 25.08}                            & 0.717     &9.342                                            \\ \hline

\rowcolor[HTML]{FFFFFF} 
\cellcolor[HTML]{EFEFEF}\textbf{MPRNet}~\cite{MPRNet}        & 29.89 & 0.904  &6.918              & 31.73                                                   & 0.905    & 6.134                                            & {\color[HTML]{333333} 25.73}                            & 0.756        & 7.653                                       \\ \hline

\rowcolor[HTML]{FFFFFF} 
\cellcolor[HTML]{EFEFEF}\textbf{TSA-Net}~\cite{TSA-net}   &30.12   &0.895   & 6.809                  & 32.41                                                   & 0.909           & 5.416                                      & {\color[HTML]{333333} 24.48}                            & 0.704    &9.642                                            \\ \hline

\rowcolor[HTML]{FFFFFF} 
\cellcolor[HTML]{EFEFEF}\textbf{MST++}~\cite{MST++}  & 30.03   &0.908  & 6.817                      & 33.57                                                   & 0.938         & 5.456                                       & {\color[HTML]{333333} 25.63}                            & 0.754       & 8.433                                        \\ \hline

\rowcolor[HTML]{FFFFFF} 
\cellcolor[HTML]{EFEFEF}\textbf{Restormer}~\cite{restormer}     &31.59  &0.929 &6.059              & \cellcolor[HTML]{DAE8FC}34.98                   & \cellcolor[HTML]{DAE8FC}0.952                 & \cellcolor[HTML]{DAE8FC}{\color[HTML]{036400} \textbf{4.688}}  & {\color[HTML]{333333} 28.95}                            & {\color[HTML]{000000} 0.847}    &\cellcolor[HTML]{DAE8FC}5.917                       \\ \hline

\rowcolor[HTML]{FFFFFF} 
\cellcolor[HTML]{EFEFEF}\textbf{CSST-9stg}~\cite{ADIS}         & \cellcolor[HTML]{DAE8FC}32.31                                                   & \cellcolor[HTML]{DAE8FC}0.939            & \cellcolor[HTML]{DAE8FC}6.014                             & 34.89                     & 0.950                                                 & 4.787                   & \cellcolor[HTML]{DAE8FC}30.10 & \cellcolor[HTML]{DAE8FC}0.864            & 6.000               \\ \hline

\rowcolor[HTML]{F1E9E1} 
\textbf{HSFAUT-3stg (Ours)}                                   & {\color[HTML]{036400} \textbf{33.60}}                   & {\color[HTML]{036400} \textbf{0.945}}                 & {\color[HTML]{036400} \textbf{5.480}}   & {\color[HTML]{036400} \textbf{35.35}}                   & {\color[HTML]{036400} \textbf{0.954}}                 & 4.732 & {\color[HTML]{036400} \textbf{31.30}}                   & {\color[HTML]{036400} \textbf{0.887}}                 & {\color[HTML]{036400} \textbf{5.621}}                                  \\ \hline

\rowcolor[HTML]{F1E9E1} 
\textbf{HSFAUT-4stg (Ours)}    & {\color[HTML]{3531FF} \textbf{33.93}}    & {\color[HTML]{3531FF} \textbf{0.949}} & {\color[HTML]{CB0000} \textbf{5.223}}                                 & {\color[HTML]{3531FF} \textbf{35.55}} & {\color[HTML]{3531FF} \textbf{0.957}} & {\color[HTML]{3531FF} \textbf{4.665}}                  & {\color[HTML]{3531FF} \textbf{32.60}}                 & {\color[HTML]{3531FF} \textbf{0.912}}                   & {\color[HTML]{3531FF} \textbf{5.416}}                \\ \hline

\rowcolor[HTML]{F1E9E1} 
\textbf{HSFAUT-5stg (Ours)}    & {\color[HTML]{CB0000} \textbf{34.14}}    & {\color[HTML]{CB0000} \textbf{0.952}} & {\color[HTML]{3531FF} \textbf{5.302}}                                 & {\color[HTML]{CB0000} \textbf{36.11}} & {\color[HTML]{CB0000} \textbf{0.962}} & {\color[HTML]{CB0000} \textbf{4.587}}                  & {\color[HTML]{CB0000} \textbf{34.04}}                 & {\color[HTML]{CB0000} \textbf{0.928}}                   & {\color[HTML]{CB0000} \textbf{4.708}}                \\ \hline


\end{tabular}
}
\end{center}
\label{tab-sdi}
\vspace{-0.4cm} 
\setlength{\belowcaptionskip}{-0.4cm}
\end{table*}

\begin{figure*}[t]
\begin{center}
  \includegraphics[width=1\linewidth]{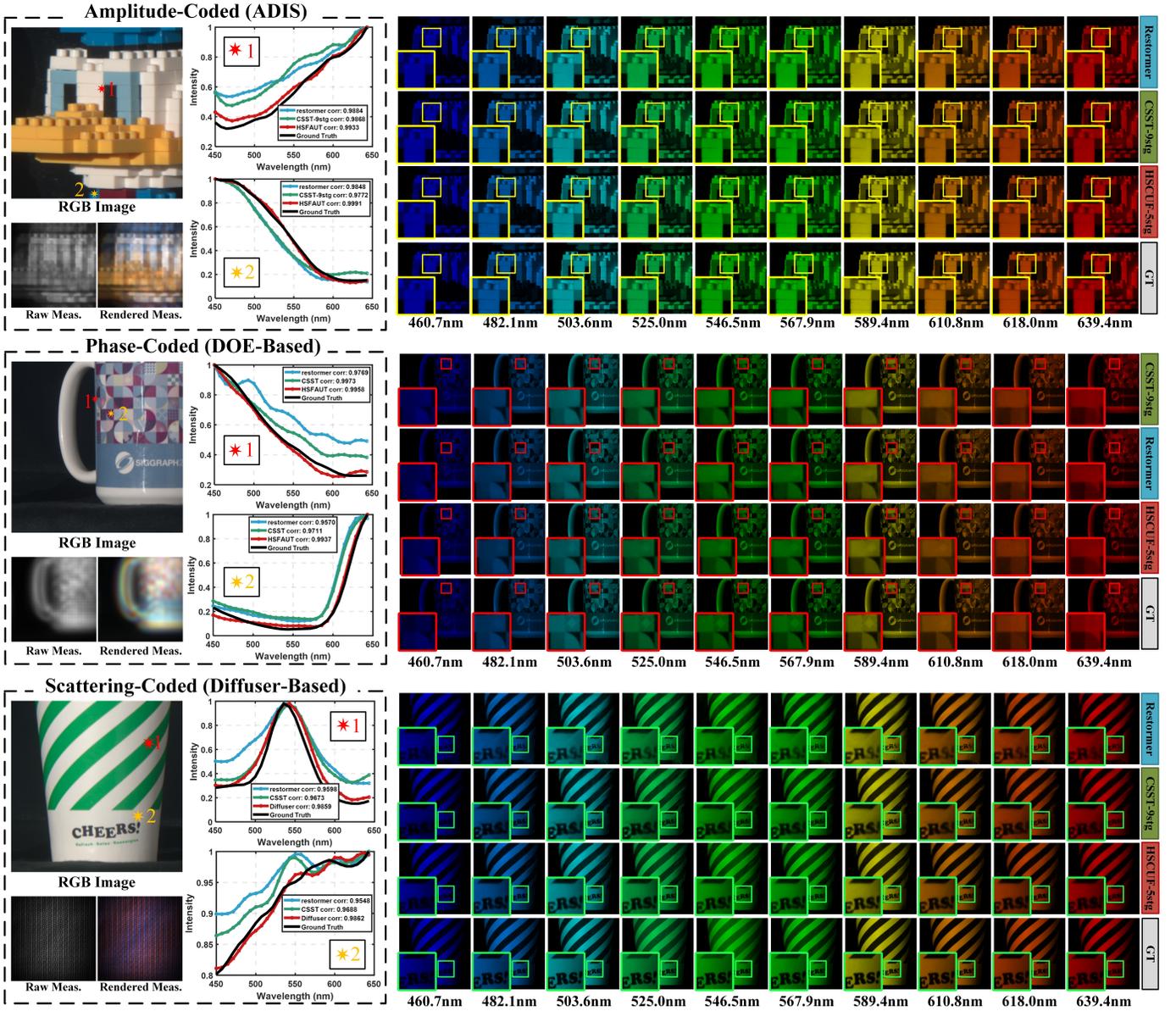}
\end{center}
\vspace{-0.2cm}
\caption{Simulation HSI reconstruction comparisons of different scenes with 10 (out of 28) spectral channels on three typical SDI systems. The left side displays the RGB image of the scene alongside the simulated measurements, while the central side presents a comparison between the reconstructed intensity-spectral curves at marked points on the RGB image and the corresponding Ground Truth. Please zoom in for a better view.}
\label{Fig5}
\vspace{-0.3cm} 
\setlength{\belowcaptionskip}{-0.3cm}
\end{figure*}

\subsection{Quantitative Comparisons}
Table~\ref{tab1} compares the reconstruction results on ADIS ( an typical amplitude coding SDI system)~\cite{ADIS} of HSFAUT and 8 SOTA methods including four super resolution (SR) algorithms (UNet~\cite{Unet}, MIRNet~\cite{MIRNet}, MPRNet~\cite{MPRNet}, TSA-Net~\cite{TSA-net}) and four spectal reconstruction methods ($\lambda$-Net~\cite{l-net}, MST++~\cite{MST++}, Restormer~\cite{restormer}, CSST-9stg) on 10 simulation scenes. All algorithms are tested with the same settings as \cite{MST,DAUHST}.
Additionally, we also conduct extensive simulation experiments on the phase-encoded DOE-based method~\cite{DOE_Jeon} and the scatter-encoded diffuser-based method~\cite{diffusercam}, with the same subjects and experimental setup as in the ADIS reconstruction, and the experimental results are shown in Table~\ref{tab-sdi}. 


The proposed model HSFAUT yields very impressive results on different SDI systems:


(i) HSFAUT-3stg achieves superior reconstruction performance across all three categories of SDI systems over previous state-of-the-art methods, including both Restormer and CSST-9stg, while utilizing only $\textbf{19.51} \%$ of the memory and $\textbf{73.65} \%$ of the computational cost of Restormer.

(ii) On the amplitude-coded SDI system, HSFAUT-5stg achieve 34.14 dB in PSNR and 0.952 in SSIM, outperform the suboptimal CSST-9stg by \textbf{1.83} dB in PSNR, \textbf{0.013} in SSIM,  \textbf{0.712} in SAM; exceeding the Restormer by \textbf{2.55} dB in PSNR. 
on the phase-coded SDI system (DOE-based), HSFAUT-5stg outperform the suboptimal Restormer by \textbf{1.13} dB in PSNR, \textbf{0.010} in SSIM, \textbf{0.101} in SAM; exceeding the CSST-9stg by \textbf{1.22} dB in PSNR; 
on the scattering-coded SDI system (diffuser-based), HSFAUT-5stg outperform the suboptimal CSST-9stg by \textbf{3.94} dB in PSNR, \textbf{0.064} in SSIM, \textbf{1.209} in SAM; exceeding the Restormer by \textbf{5.09} dB in PSNR. 

(iii) Our HSFAUT models dramatically surpass SOTA methods while requiring cheaper memory and computational costs. Even when the number of iterations increased to five, we still achieve a significant surpass on all three encoded SDI systems with only \textbf{59.75$\%$} of the Params costs of CSST-9stg and \textbf{25.92$\%$} of the Params costs of Restormer. 


Building on the above analysis, we further observe that the absolute improvement offered by HSFAUT is closely linked to the degree of spatial degradation: systems with more severe degradation exhibit greater potential for performance gains. On the severely degraded scattering-encoding Spectral DiffuserCam~\cite{diffusercam}, HSFAUT-5stg achieves an improvement exceeding \textbf{5} dB with only 25.92$\%$ memory overhead of Restormer. This result strongly suggests that HSFAUT can effectively captures critical cues that degraded in the spatial domain through spatial frequency aggregation, thereby significantly enhancing reconstruction quality.

Additionally, achieving optimal performance across diverse SDI systems remains challenging, as different encoding principles produce PSFs with distinct spatial characteristics. For instance, phase encoding (e.g., using DOEs) yields PSFs with relatively concentrated energy distribution, while amplitude and scattering-based encoding often result in widely dispersed projection patterns. These differences impose varying requirements on the effective receptive field of reconstruction algorithms. As evidenced empirically, Restormer outperforms CSST-9stg on DOE-based phase-encoded systems, whereas CSST-9stg excels in amplitude- and scattering-encoded systems due to its architectural design. In contrast, our proposed HSFAUT leverages frequency-domain properties common to all SDI systems, enabling consistent and superior performance across different SDI encoding principles.

\subsection{Qualitative Comparisons}

\subsubsection{Simulation HSIs Reconstruction}

Figure \ref{Fig5} compares the HSIs reconstructions of our HSFAUT-5stg method with those of other leading  algorithms across various SDI systems. The bottom-left inset displays zoomed-in patches from the full HSI.
It can be observed that our HSFAUT-5stg consistently produces HSIs with superior visual quality, exhibiting clearer textures and enhanced image details that closely resemble the ground truth (GT). In contrast, prior methods exhibit two distinct types of shortcomings: they either yield over-smoothed results that compromise fine structural details, as seen in ADIS and DOE-based systems, or introduce chromatic artifacts and speckle textures absent in the ground truth, as observed in Diffuser-based systems. 

Furthermore, the intensity-wavelength spectral profiles at the heptagram location marked in the RGB image (left) further substantiate our method's efficacy. The profiles obtained by HSFAUT-5stg show the strongest agreement with the reference, underscoring the advantages of our proposed framework in achieving spectrally consistent reconstruction.

\subsubsection{Visualization of the HSFAUT}
\begin{figure*}[h]
    \begin{center}
    \includegraphics[width=0.98\linewidth]{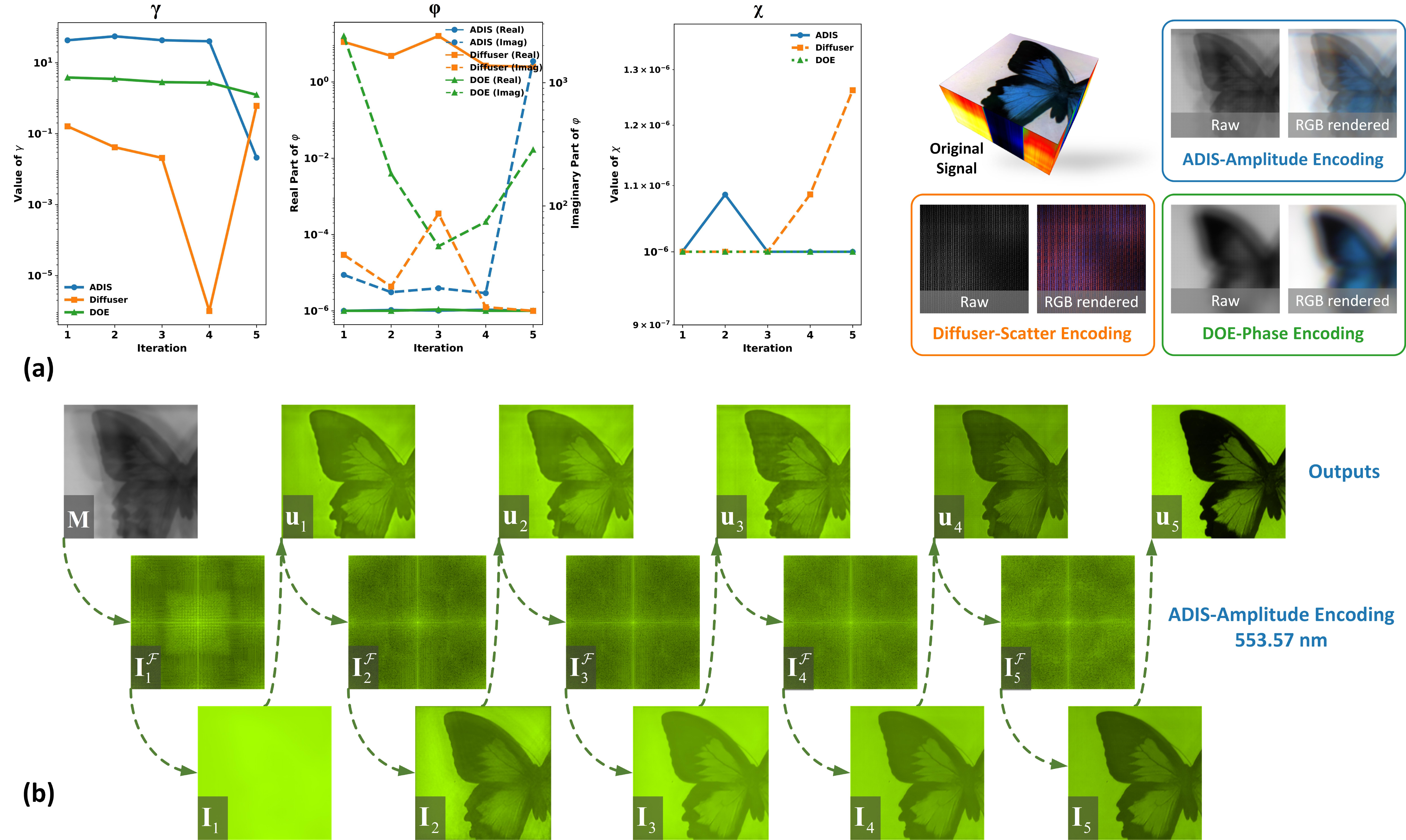}
    \end{center}
\vspace{-0.2cm} 
\caption{Visualisation of parameter variations (a) and intermediate variable variations (b) of HSFAUT-5stg. More visualisation of intermediate variable variations, please refer to \textbf{supplementary materials}.}
\vspace{-0.2cm} 
\label{Fig-param}
\end{figure*}

To further analyze the roles of the estimated parameters in different SDI systems, we plot the curves of estimated parameters and visualize intermediate variables as they change with the iteration in Fig.~\ref{Fig-param}.
We observe: 

(1) The parameters $\boldsymbol{\gamma}$ and $\boldsymbol{\varphi}$, which control the linear update in the filtering and convolutional subproblems respectively, exhibit a consistent relationship with the degree of spatial blur in different SDI systems. Specifically, stronger degradation correlates with smaller values of $\boldsymbol{\gamma}$ and a smaller real part of $\boldsymbol{\varphi}$. 

(2) In the convolutional subproblem, the frequency-domain update involves taking the real part of complex-valued intermediates. As a result, the parameter $\boldsymbol{\chi}$, which is related to noise-level, may occasionally increase abruptly during iterations to suppress artifacts—a behavior that deviates from the conventional monotonic decrease observed in deep unfolding methods~\cite{DAUHST}.

\subsection{Ablation Study}

\begin{table}[t]

\caption{Ablation experiments conducted on the frequency-spectral branch (FS branch) of the SFAT.}
\vspace{-0.3cm}
\begin{center}
\renewcommand{\arraystretch}{1.2} 
\setlength{\tabcolsep}{3.2pt}     
\resizebox{0.98\columnwidth}{!}{
\begin{tabular}{c|c|cc|cc}
\hline
\rowcolor[HTML]{EFEFEF} 
\textbf{FS Branch} & \textbf{Encoding} & \textbf{PSNR} & \textbf{SSIM} & \textbf{Params}               & \textbf{GFLOPs}               \\ \hline
\ding{55}                 & Amplitude         & 32.84             & 0.941             & 2.91M                             & 64.30                             \\
\rowcolor[HTML]{EFEFEF} 
\checkmark                  & Amplitude         & \bf{33.60}             & \bf{0.945}             & \cellcolor[HTML]{EFEFEF}2.95M & \cellcolor[HTML]{EFEFEF}64.72 \\ \hline
\ding{55}                  & Phase             & 35.06             & 0.953             & 2.91M                             & 64.30                             \\
\rowcolor[HTML]{EFEFEF} 
\checkmark                  & Phase             & \bf{35.35}             & \bf{0.954}             & 2.95M                         & 64.72                         \\ \hline
\ding{55}                  & Scatter           & 30.24             & 0.869             & 2.91M                             & 64.30                             \\
\rowcolor[HTML]{EFEFEF} 
\checkmark                  & Scatter           & \bf{31.30}             & \bf{0.887}             & 2.95M                         & 64.72                         \\ \hline
\end{tabular}
}
\end{center}
\label{tab-SFAT}
\vspace{-0.3cm} 
\end{table}

\begin{table}[t]
\caption{Ablation on architecture of HSFAUF, PSNR (dB), SSIM, params and GFLOPs are reported.}
\vspace{-0.2cm}
\begin{center}
\renewcommand{\arraystretch}{1.2} 
\setlength{\tabcolsep}{3.2pt}     
\resizebox{0.98\columnwidth}{!}{
\begin{tabular}{cc|cc|c|cc|cc}
\hline
\rowcolor[HTML]{EFEFEF} 
\multicolumn{2}{c|}{\cellcolor[HTML]{EFEFEF}\textbf{Filtering}} & \multicolumn{2}{c|}{\cellcolor[HTML]{EFEFEF}\textbf{Conv}} & \cellcolor[HTML]{EFEFEF}                                    & \cellcolor[HTML]{EFEFEF}                                & \cellcolor[HTML]{EFEFEF}                                & \cellcolor[HTML]{EFEFEF}                                  & \cellcolor[HTML]{EFEFEF}                                  \\
\rowcolor[HTML]{EFEFEF} 
$\boldsymbol{\gamma}$               & $\cal U$               & $\boldsymbol{\varphi}$     & $\boldsymbol{\chi}$    & \multirow{-2}{*}{\cellcolor[HTML]{EFEFEF}\textbf{Encoding}} & \multirow{-2}{*}{\cellcolor[HTML]{EFEFEF}\textbf{PSNR}} & \multirow{-2}{*}{\cellcolor[HTML]{EFEFEF}\textbf{SSIM}} & \multirow{-2}{*}{\cellcolor[HTML]{EFEFEF}\textbf{Params}} & \multirow{-2}{*}{\cellcolor[HTML]{EFEFEF}\textbf{GFLOPs}} \\ \hline
\ding{55}         & \ding{55}       & \ding{55}              & \ding{55}           & Amplitude                                                   & 29.95                                                       & 0.909                                                      & 2.95M                                                         & 18.43                                                         \\
\ding{55}        & \ding{55}      & \checkmark          & \checkmark      & Amplitude                                                   & 33.08                                                      & 0.943                                                       & 2.95M                                                         & 55.37                                                         \\
\checkmark           & \checkmark          & \ding{55}     & \ding{55}      & Amplitude                                                   & 30.06                                                       & 0.914                                                      & 2.95M                                                        & 18.43                                                         \\
\rowcolor[HTML]{EFEFEF} 
\checkmark                  & \checkmark     & \checkmark       & \checkmark   & Amplitude                                                   & 33.60                                                       & 0.945                                                       & 2.95M                                                     & 64.72                                                     \\ \hline
\ding{55}           & \ding{55}      & \ding{55}        & \ding{55}             & Phase                                              & 33.88                                                       & 0.940                                                      & 2.95M                         & 18.43                                                         \\
\rowcolor[HTML]{EFEFEF} 
\checkmark        & \checkmark       & \checkmark          & \checkmark                 & Phase                                                       & 35.35                                                       & 0.954                                                     & 2.95M                                                     & 64.72                                                     \\ \hline
\ding{55}       & \ding{55}  & \ding{55}      & \ding{55}       & Scatter                                                     & 28.20                                                      & 0.809                                                      & 2.95M                                          & 18.43                                                         \\
\rowcolor[HTML]{EFEFEF} 
\checkmark     & \checkmark     & \checkmark      & \checkmark    & Scatter                                                     & 31.30                                                      & 0.887                                                      & 2.95M                                                     & 64.72                                                     \\ \hline
\end{tabular}
}
\end{center}
\label{tab-HSFAUF}
\vspace{-0.4cm} 
\setlength{\belowcaptionskip}{-0.4cm}
\end{table}

\subsubsection{Ablation on SFAT}
In SFAT, we introduce a distinctive architecture that processes frequency-domain amplitude features and integrates them with spatial-spectral multi-head attention outputs via a learnable gating mechanism. This design explicitly aggregates discriminative information from both domains. To validate the necessity of this frequency–spectral branch as a core component of the denoiser, we conduct ablation studies across various SDI systems. Using HSFAUT-3stg as the baseline, comparative results (Table~\ref{tab-SFAT}) under the same experimental setup as Section~\ref{subsec:Setup} demonstrate that the proposed branch significantly enhances reconstruction quality with minimal additional computational and memory overhead. These findings confirm both the effectiveness of the frequency–spectral feature aggregation and the overall soundness of SFAT’s design.


\subsubsection{Ablation on HSFAUF}

We further conduct an ablation study to evaluate the contributions of key components in the proposed HSFAUF, and we use HSFAUT-3stg to conduct the experiments in order to investigate the impact of each component on higher performance. All models in this study were trained using the same configuration detailed in Section~\ref{subsec:Setup}.


As summarized in Table~\ref{tab-HSFAUF}, incorporating the bootstrap parameters for the filtering and convolutional subproblems improves PSNR on the ADIS test set by 0.52 dB and 3.54 dB, respectively. Their combined use yields a further gain, resulting in a total improvement of 3.65 dB. Similar significant gains are observed for phase-encoding (1.47 dB) and scattering-encoding (3.10 dB) SDI tasks. These results demonstrate that the performance of HSFAUF stems primarily from its carefully orchestrated frequency-domain updates, which provide explicit guidance to the denoiser and lead to substantial improvements in final reconstruction quality.



\subsubsection{Comparison on iteration numbers}

Table~\ref{tab-iter} compares the reconstruction performance across different iteration counts. Both computational and memory costs scale linearly with the number of stages. The HSFAUT-5stg configuration achieves near-optimal performance, representing a favorable trade-off between efficiency and reconstruction quality.

For both phase-encoding(DOE-based) and scatter-encoding (diffuser-based) systems, increasing the number of stages improves reconstruction quality, albeit with diminishing returns beyond five stages. Consequently, HSFAUT-5stg is identified as a practical operating point that balances high performance with manageable computational expense.

\begin{table}[t]
\caption{Quantitative comparison on iteration numbers, PSNR (dB), SSIM, params and GFLOPs are reported.}
\vspace{-0.3cm}
\begin{center}
\renewcommand{\arraystretch}{1.25} 
\setlength{\tabcolsep}{3.3pt}     
\resizebox{1.0\columnwidth}{!}{
\begin{tabular}{c|cc|cc|cc|cc}
\hline
\rowcolor[HTML]{EFEFEF} 
\cellcolor[HTML]{EFEFEF}                            & \multicolumn{2}{c|}{\cellcolor[HTML]{EFEFEF}\textbf{Amplitude}} & \multicolumn{2}{c|}{\cellcolor[HTML]{EFEFEF}\textbf{Phase}} & \multicolumn{2}{c|}{\cellcolor[HTML]{EFEFEF}\textbf{Scatter}} & \cellcolor[HTML]{EFEFEF}                                  & \cellcolor[HTML]{EFEFEF}                                  \\
\rowcolor[HTML]{EFEFEF} 
\multirow{-2}{*}{\cellcolor[HTML]{EFEFEF}\bf{Algorithm}} & \textbf{PSNR}                       & \textbf{SSIM}                      & \textbf{PSNR}                     & \textbf{SSIM}                    & \textbf{PSNR}                      & \textbf{SSIM}                     & \multirow{-2}{*}{\cellcolor[HTML]{EFEFEF}\textbf{Params}} & \multirow{-2}{*}{\cellcolor[HTML]{EFEFEF}\textbf{GFLOPs}} \\ \hline
\cellcolor[HTML]{EFEFEF}\textbf{Ours-2stg}        & 32.65                               & 0.938                              & 34.36                             & 0.944                            & 28.34                              & 0.827                             & 2.46M                                                     & 46.71                                                     \\
\cellcolor[HTML]{EFEFEF}\textbf{Ours-3stg}        & 33.60                               & 0.945                              & 35.35                             & 0.954                            & 31.30                              & 0.887                             & 2.95M                                                     & 64.72                                                     \\
\cellcolor[HTML]{EFEFEF}\textbf{Ours-4stg}        & 33.93                               & 0.949                              & 35.55                             & 0.957                            & 32.60                              & 0.912                             & 3.43M                                                     & 82.72                                                     \\
\cellcolor[HTML]{EFEFEF}\textbf{Ours-5stg}        & 34.14                               & 0.952                              & 36.11                             & 0.962                            & 34.04                              & 0.928                             & 3.92M                                                     & 100.72                                                    \\
\cellcolor[HTML]{EFEFEF}\textbf{Ours-6stg}        & 33.94                               & 0.953                              & 36.31                             & 0.965                            & 34.39                              & 0.934                             & 4.41M                                                     & 118.72                                                    \\ \hline
\end{tabular}
}
\end{center}
\label{tab-iter}
\vspace{-0.3cm} 
\setlength{\belowcaptionskip}{-0.4cm}
\end{table}

\section{Real Experiment}
\label{sec:real}

\begin{figure*}[t]
  \centering
  \includegraphics[width=1.0\linewidth]{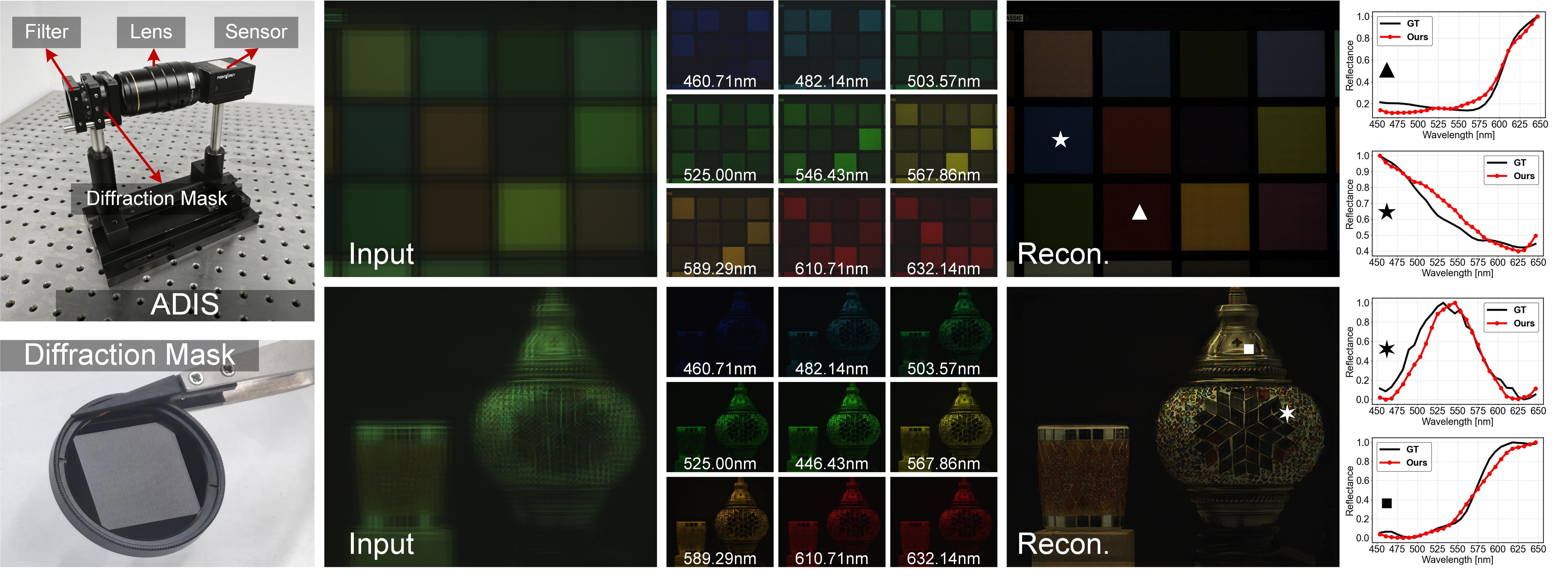}
   \vspace{-0.4cm}
   \caption{We built ADIS prototype  to perform the real captures, with parameters consistent with \cite{ADIS}. HSFAUT is able to faithfully recover the spectral and spatial details in real acquisitions.}
   \vspace{-0.2cm}
   \label{Fig-real}
\end{figure*}

\subsection{Real Experiment Setup}
\label{subsec:RealSetup}
We conducted \textbf{real experiments} by developing prototype of ADIS~\cite{ADIS} as demonstrated in Fig~\ref{Fig-real}(a). 
The ADIS prototype utilizes a binary mask as its core encoding element, an RGB industrial camera (FLIR GS3-U3-51S5C) for signal capture, and a bandpass filter (450–650 nm) to define the operational spectral range. The mask is fabricated with two layers of 100-µm-wide opaque parallel lines crossed orthogonally, featuring a line-width uniformity of $0.2 \mu m$ and a spacing equal to the line width.
Using these hardware parameters, we computed the wavelength-dependent PSFs of the ADIS and established a forward model for measurement synthesis.
 Under this experimental setup, HSFAUT-3stg was trained jointly on the CAVE-1024 and KAIST datasets, with its effectiveness subsequently validated through real-world captures.


\subsection{Real Experiment Results}
\label{subsec:RealResults}

Real-world HSIs reconstruction results are presented in Fig. \ref{Fig-real}(b) and (c). Our method accurately reconstructs the captured scenes, recovering fine textures and spatial details. The resulting spectral profiles show a close match to point spectrometer measurements, producing structurally coherent images free of artifacts.

As outlined in Section~\ref{sec:intro}, SDI achieves high spectral fidelity within a compact optical design—a capability clearly reflected in these real-world experiments. In contrast, array-pattern encoding (APE) methods, which rely solely on transmittance or response encoding, often exhibit limited fidelity. Meanwhile, integral-path modulation (IPM) approaches tend to require more complex optical configurations.

\section{Discussion}
%

\subsection{The Generalisability of HSFAUF}
HSFAUF, as the core contribution of this work, leverages the proposed SFAT to achieve consistently superior performance across all three categories of SDI reconstruction tasks. However, It is important to note that HSFAUF is a flexible deep unfolding framework that can integrate various denoisers to further enhance reconstruction quality.  To systematically evaluate the efficiency and generalization capability of HSFAUF for SDI inverse problems, we embed several recently proposed advanced denoisers originally designed for CASSI reconstruction into our framework, and compare their performance against a simple cascade of multiple denoisers.


\begin{table}[h]
\caption{Quantitative comparison of cascade denoisers and denoisers embedded in HSFAUF.}
\vspace{-0.2cm}
\begin{center}
\renewcommand{\arraystretch}{1.25} 
\setlength{\tabcolsep}{3.3pt}     
\resizebox{1.0\columnwidth}{!}{
\begin{tabular}{c|cc|cc}
\hline
\rowcolor[HTML]{EFEFEF} 
\textbf{Algorithms}                  & \textbf{PSNR} & \textbf{SSIM}  & \textbf{Params} & \textbf{GFLOPs} \\ \hline
Cascade-UNet~\cite{Unet}    & 31.03    & 0.905      & 69.98M     & 16.15      \\
\rowcolor[HTML]{EFEFEF} 
HSFAUF-UNet                 & 32.52    & 0.934       & 71.57M      & 64.98      \\
Cascade-Trident-Transformer~\cite{wu2024latent} & 30.61    & 0.916       & 1.83M      & 23.93      \\
\rowcolor[HTML]{EFEFEF} 
HSFAUF-Trident-Transformer  & 32.83    & 0.943       & 3.42M      & 72.77      \\
Cascade-Mix$S^2$ Transformer~\cite{dong2023residual}                & 29.53    & 0.906       & 1.88M      & 28.95      \\
\rowcolor[HTML]{EFEFEF} 
HSFAUF-Mix$S^2$ Transformerr                     & 32.98    & 0.947       & 3.48M      & 77.78      \\ \hline
\end{tabular}
}
\end{center}
\label{tab-ge}
\vspace{-0.3cm} 
\setlength{\belowcaptionskip}{-0.4cm}
\end{table}

Here, the classical UNet \cite{Unet}, the Trident-Transformer from \cite{wu2024latent} (ECCV2024), and the Mix$S^2$ Transformer from RDLUF-Mix$S^2$ \cite{dong2023residual} (CVPR2023) is selected for comparison. All evaluations use a 3-stage HSFAUF configuration, with the baseline being a direct cascade of three identical denoisers. 
As shown in Table \ref{tab-ge}, comparative experiments conducted on the ADIS (amplitude-encoding) demonstrate that HSFAUF consistently enhances reconstruction quality across different denoiser architectures. By effectively leveraging its spatial-frequency aggregation mechanism, HSFAUF shows general applicability and improved performance regardless of the specific denoiser employed.

\subsection{Frequency-to-Spatial Conversion Strategies}
\label{sec6.2}
 The complex division in Eq.\ \ref{eq.3.2.17} disrupts Hermitian symmetry, rendering amplitude-based frequency-to-spatial conversion suboptimal (Section~\ref{subsec:HSFAUF}). Although direct denoising in the frequency domain would avoid this issue, it introduces prohibitive computational and memory costs, which initially impeded our progress. However, replacing the amplitude extraction step with taking the real part of the complex values led to a marked improvement in reconstruction quality. Although the imaginary component of the complex field is ultimately discarded after the linear update in the convolutional subproblem, it appears to serve as an implicit computational space that supports more stable and informative transformations.
Here we compare different strategies for converting frequency-domain back to the spatial domain, evaluating HSFAUT-3stg on the ADIS. Here, ${\cal A} ( \cdot )$ denotes using the amplitude, $\Re ( \cdot )$ the real part, and ${\cal I} ( \cdot )$ the imaginary part. Quantitative results are summarized in Table \ref{tab-transform}.

\begin{table}[h]
\vspace{-0.1cm}
\caption{Quantitative comparison of different strategies for converting frequency domain back to the spatial domain.}
\vspace{-0.3cm}
\begin{center}
\renewcommand{\arraystretch}{1.25} 
\setlength{\tabcolsep}{3.3pt} 
\resizebox{0.9\columnwidth}{!}{
\begin{tabular}{cc|cc}
\hline
\rowcolor[HTML]{EFEFEF} 
\textbf{Algorithm}   & \textbf{Transforming Method} & \textbf{PSNR} & \multicolumn{1}{c}{\cellcolor[HTML]{EFEFEF}\textbf{SSIM}} \\ \hline
HSFAUT-3stg & ${\cal A} ( \cdot )$                            & \textbf{32.54}         & \textbf{0.940}                                                      \\
HSFAUT-3stg & ${\cal I} ( \cdot )$                            & 33.09         & 0.941                                                      \\
HSFAUT-3stg & $\Re ( \cdot )$                            & \textbf{33.60}         & \textbf{0.945}                                                      \\ \hline
\end{tabular}
}
\end{center}
\label{tab-transform}
\vspace{-0.3cm} 
\setlength{\belowcaptionskip}{-0.4cm}
\end{table}

The results clearly demonstrate that under nearly identical computational and memory constraints, $\Re ( \cdot )$ achieves superior performance compared to both amplitude and imaginary part extraction. Thus, while taking the real part may appear as a straightforward implementation detail, it proves essential to HSFAUF's effectiveness.

\subsection{Noise Analysis}
\label{sec6.3}
In practical imaging scenarios, measurements are often degraded by various noise sources, such as photon shot noise, read noise, and dark current, causing the acquired data to deviate from the ideal forward model. To systematically evaluate the impact of noise on reconstruction performance across different SDI systems and algorithms, we introduce Gaussian noise into representative SDI systems for a controlled comparative study. The experimental results, summarized in Table~\ref{tab-noise}, reveal how noise influences both system robustness and algorithmic stability.

\begin{table}[h]
\vspace{-0.1cm}
\caption{Quantitative comparison of reconstruction performance of different algorithm across different SDI systems under Gaussian noise degradation.}
\vspace{-0.3cm}
\begin{center}
\renewcommand{\arraystretch}{1.25} 
\setlength{\tabcolsep}{3.3pt} 
\resizebox{0.9\columnwidth}{!}{
\begin{tabular}{ccc|cc}
\hline
\rowcolor[HTML]{EFEFEF} 
\textbf{System} & \textbf{Algorithm} & $\sigma$ & \textbf{PSNR} & \multicolumn{1}{c|}{\cellcolor[HTML]{EFEFEF}\textbf{SSIM}} \\ \hline
Amplitude-Encoding       & HSFAUT-3stg        & 0.01       & \textbf{31.72}         & \textbf{0.906}                                                      \\
Amplitude-Encoding       & Restormer~\cite{restormer}          & 0.01       & 30.44         & 0.889                                                      \\
Amplitude-Encoding       & CSST-9stg~\cite{ADIS}          & 0.01       & 31.11         & 0.899                                                      \\ \hline
Phase-Encoding           & HSFAUT-3stg        & 0.01       & \textbf{28.64}               &  \textbf{0.820}                                                          \\
Phase-Encoding           & Restormer~\cite{restormer}          & 0.01       & 28.43              &   0.817                                                         \\
Phase-Encoding           & CSST-9stg~\cite{ADIS}          & 0.01       &   28.48            &   0.816                                                         \\ \hline
Scatter-Encoding         & HSFAUT-3stg        & 0.01       &  \textbf{25.36}    &  \textbf{0.731}        \\
Scatter-Encoding         & Restormer~\cite{restormer}          & 0.01       &  24.58               &  0.717                                                          \\
Scatter-Encoding         & CSST-9stg~\cite{ADIS}          & 0.01       &  24.78               &    0.721                                                        \\ \hline
\end{tabular}
}
\end{center}
\label{tab-noise}
\vspace{-0.3cm} 
\setlength{\belowcaptionskip}{-0.4cm}
\end{table}

When comparing the noise-robustness of the same algorithm across different SDI systems, phase-encoding and amplitude-encoding methods exhibit less influence under noise compared to scatter-encoding systems. This can be attributed to their relatively confined PSF energy spread, which preserves partial structural information in ideal measurements. In contrast, scatter-encoding produces highly degraded measurements where structural information is substantially lost, making scene recovery significantly more challenging under noisy conditions.

When evaluating different algorithms on the same SDI system under controlled noise degradation, HSFAUT maintains superior reconstruction quality under same noise levels. This consistent performance advantage demonstrates the HSFAUT's robustness and practical applicability for real-world SDI reconstruction tasks.

\section{Potential Research Directions}
1) Extensibility of HSFAUT:
Although our work focuses specifically on spectral deconvolution imaging (SDI), many computational imaging tasks or computer vision tasks share structurally similar forward models. For instance, super-resolution can be formulated as a wavelength-wise convolution followed by spatial downsampling, while motion deblurring corresponds to a uniform convolution across spectral bands followed by sampling with a monochromatic or color-filtered camera. Given these conceptual parallels, we believe that HSFAUT possesses general applicability to a broad class of imaging inverse problems, including but not limited to super-resolution, motion deblurring, microscopy deconvolution, astronomical image restoration, and wavefront or phase retrieval. Validating and extending the core spatial–frequency aggregation principle of HSFAUT across these domains represents a compelling direction for future work.

2) Computational Efficiency: 
Our proposed HSFAUT exhibits a clear advantage in terms of parameter efficiency; however, its computational efficiency is relatively less pronounced. This limitation stems from the iterative computations inherently required by the optimization-based training pathway of deep unfolding frameworks. For large-scale edge applications, a key challenge for compact yet high-fidelity computational imaging systems lies in developing reconstruction algorithms that are both efficient and accurate. 

\section{Conclusion}
\label{sec:conclusion}

This work presents a taxonomy of CSI methods and classifies them into three categories: IPM, APE, and SDI, according to their encoding principles and inverse problem formulations. We underscore SDI's advantages in system compactness and high fidelity but identify a critical shortcoming in existing reconstruction algorithms: their insufficient exploitation of the underlying SDI physics. The key challenge, therefore, lies in efficiently leveraging the physical priors of SDI to design more effective reconstruction algorithms. 
To address this, we reformulate the inverse problem via stepwise hierarchical solving and frequency-domain diagonalization. We propose the HSFAUF based on hierarchical spatial–frequency solving, which integrates a novel SFAT as a denoiser to form HSFAUT, thereby enhancing joint spatial–spectral feature perception. HSFAUT achieves markedly improved reconstruction accuracy while maintaining advantages in computational and memory efficiency.
Its effectiveness has been validated on real measurements using a prototype system.
Beyond SDI, HSFAUT holds strong potential for a broad range of imaging inverse and restoration problems involving convolutional encoding or degradation. Furthermore, it paves the way for joint, physics-driven optimization of lenses and filters or sensors with deep optics, fully harnessing the intrinsic advantages of SDI in system compactness and imaging fidelity.



\section*{Acknowledgments}
This research was supported by National Key Research and Development Program of China (2023YFF0713300).



 



\bibliographystyle{unsrt}
\bibliography{SDI}


\newpage

\section{Biography Section}
 



\begin{IEEEbiography}[{\includegraphics[width=1in,height=1.25in,clip,keepaspectratio]{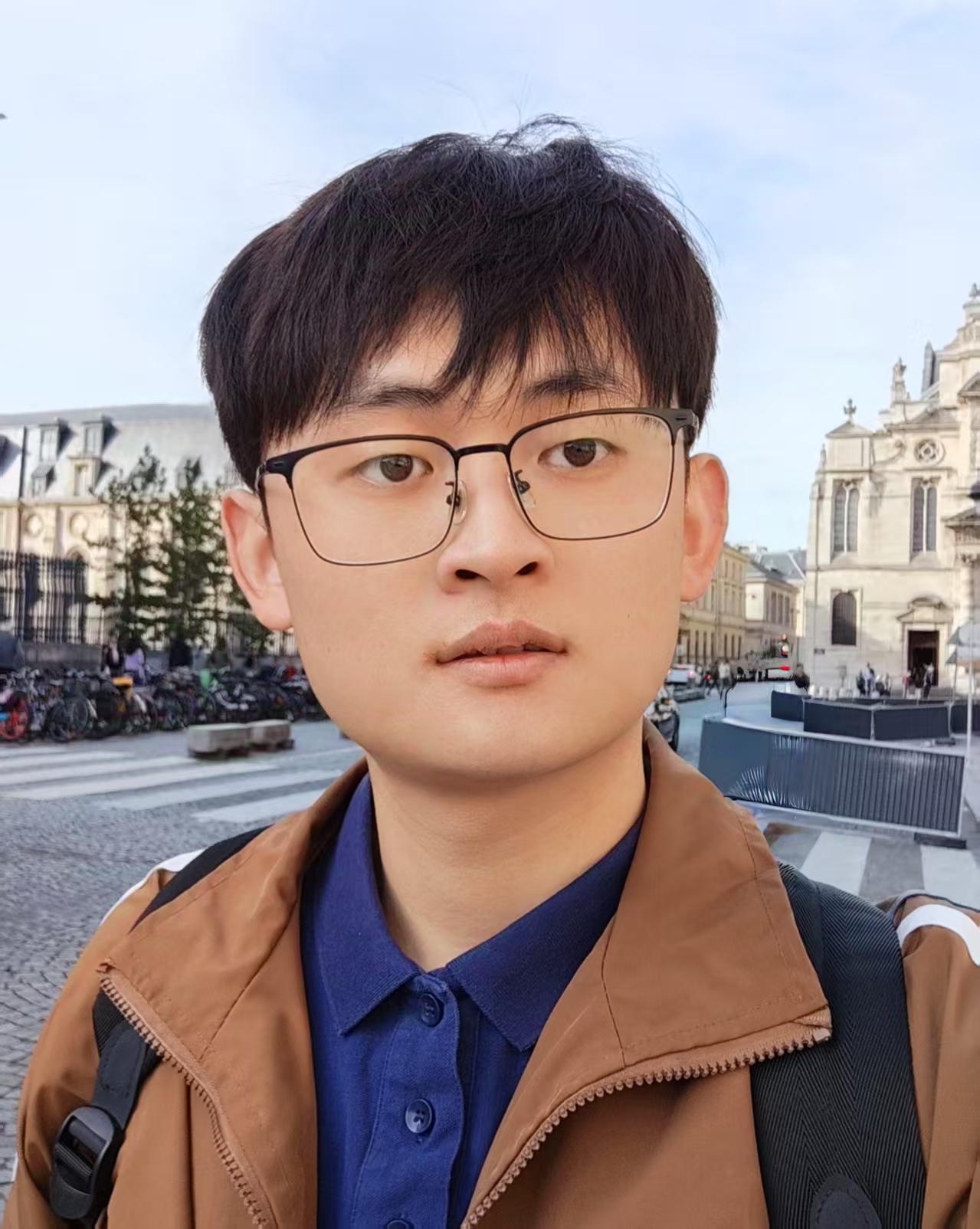}}]{Tao Lv} received the BS degree from the College of Information Science and Engineering, Northeastern University, Liaoning, China, in 2021. He is currently working toward the PhD degree with the School of Electronic Science and Engineering, Nanjing University, Nanjing, China. His research interests include computational photography and computer vision, especially computational spectral imaging.
\end{IEEEbiography}
\vspace{11pt}

\begin{IEEEbiography}[{\includegraphics[width=1in,height=1.25in,clip,keepaspectratio]{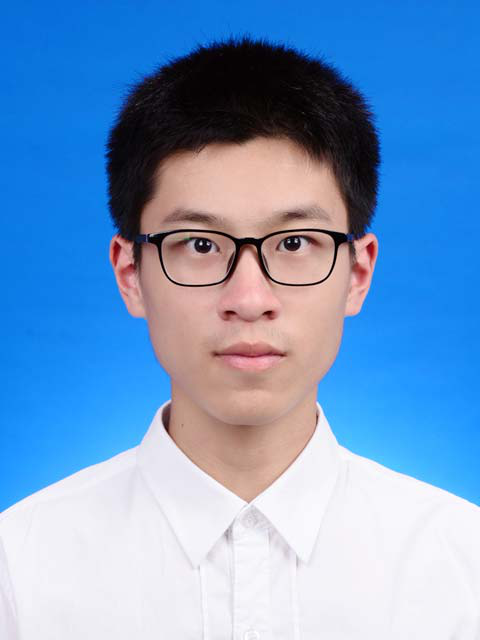}}]{Daoming Zhou} received the BS degree in optoelectronic from Nanjing University Nanjing,China, in 2025. He is currently working toward the MS degree with the School of Electronic Science and Engineering, Nanjing University. His research interests include computational imaging and computer vision.
\end{IEEEbiography}
\vspace{11pt}


\begin{IEEEbiography}[{\includegraphics[width=1in,height=1.25in,clip,keepaspectratio]{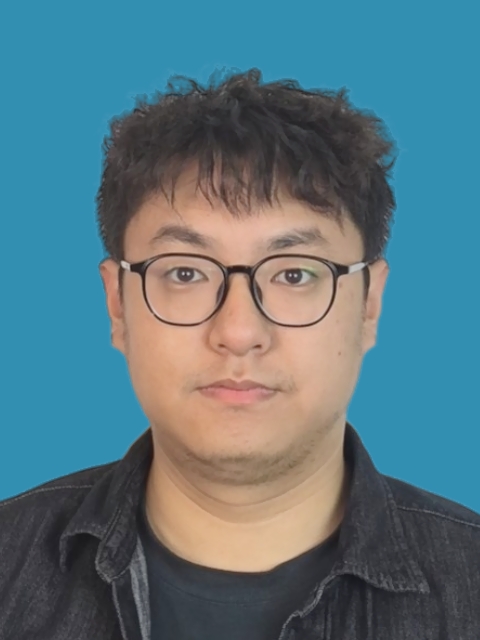}}]{Chenglong Huang} received the BS degree in communication engineering from Hohai University, China, in 2023. He is a graduate student from the School of Electronic Science and Engineering, Nanjing University.
\end{IEEEbiography}
\vspace{11pt}


\begin{IEEEbiography}[{\includegraphics[width=1in,height=1.25in,clip,keepaspectratio]{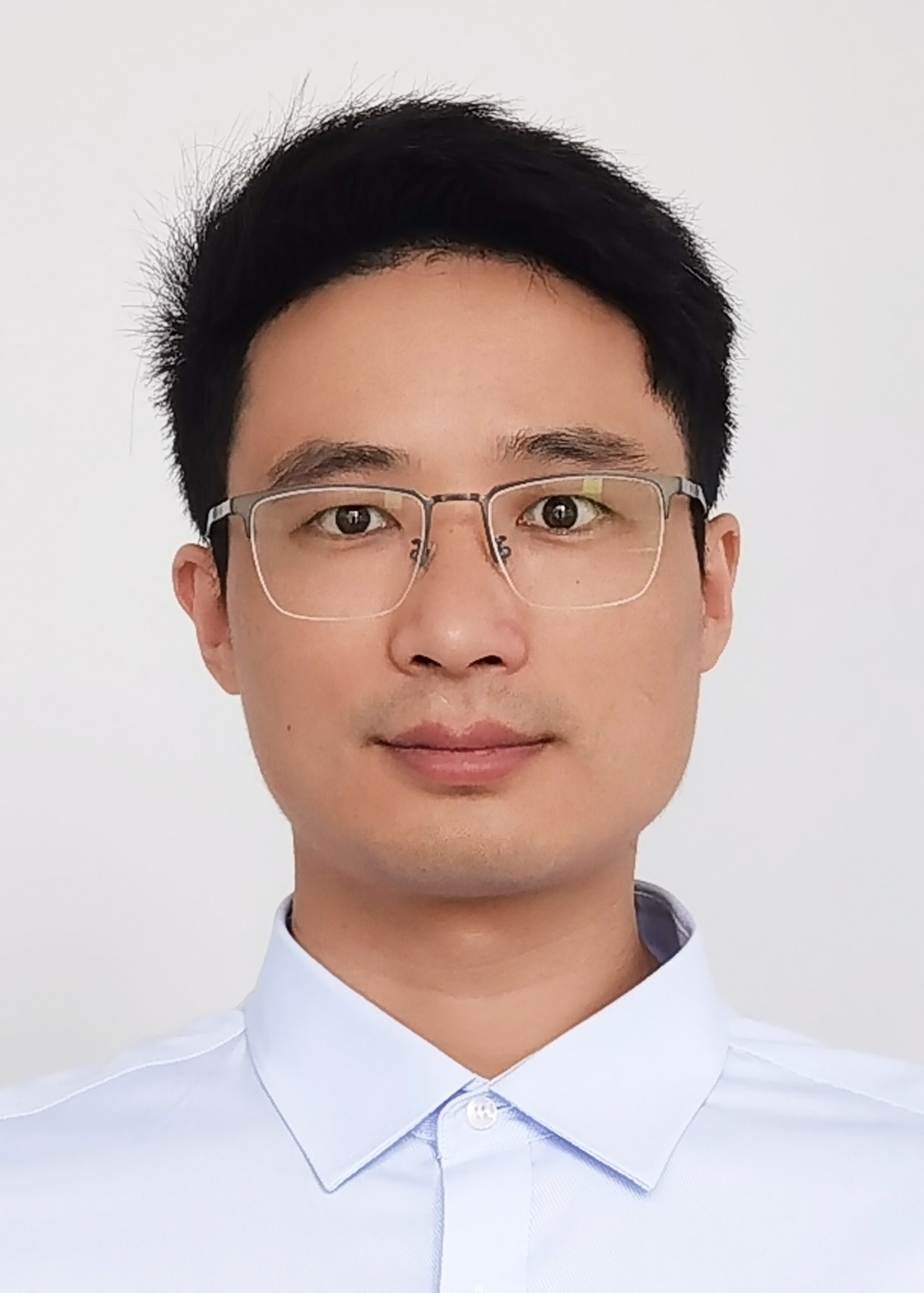}}]{Chongde Zi} received BS and MS degrees from Yunnan Normal University, Kunming, China, in 2014 and 2017, respectively, and PhD degree from Nanjing University, Nanjing, China, in 2024. He is currently a assistant researcher at Nanjing University, Nanjing, China. His research interests include computational photography and spectral imaging.
\end{IEEEbiography}
\vspace{11pt}

\begin{IEEEbiography}[{\includegraphics[width=1in,height=1.25in,clip,keepaspectratio]{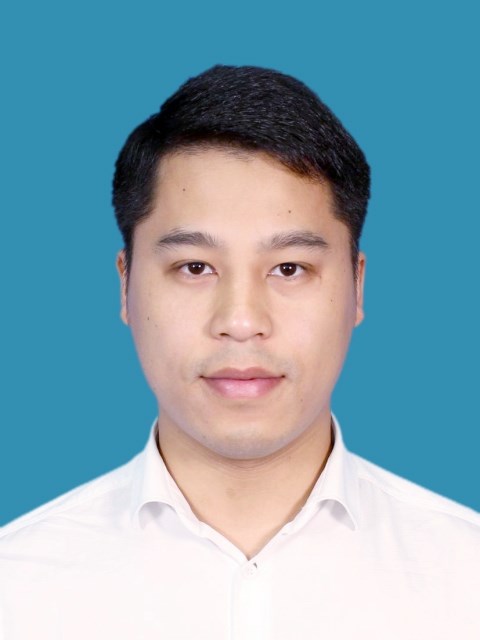}}]{Linsen Chen}
received the BS, MS and PhD degrees in 2014, 2017 and 2023 from School of Electronic Science and Engineering, Nanjing University, Nanjing, China. His research interests include computational photography, spectral imaging and reconstruction.
\end{IEEEbiography}
\vspace{11pt}


\begin{IEEEbiography}[{\includegraphics[width=1in,height=1.25in,clip,keepaspectratio]{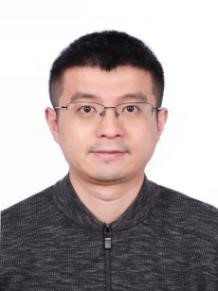}}]{Xun Cao}
(Member, IEEE) received the BS degree from Nanjing University, Nanjing, China, in 2006, and the PhD degree from the Department of Automation, Tsinghua University, Beijing, China, in 2012. He held visiting positions with Philips Research, Aachen, Germany, in 2008, and Microsoft Research Asia, Beijing, from 2009 to 2010. He was a visiting scholar with The University of Texas at Austin, Austin, Texas, from 2010 to 2011. He is currently a professor with the School of Electronic Science and Engineering, Nanjing University. His research interests include computational photography, image-based modeling and rendering, and VR/AR systems.
\end{IEEEbiography}
\vspace{11pt}


\vfill

\end{document}